\documentclass[a4paper, 11pt]{article}
\pdfoutput=1
\usepackage{aapreprint}
\makeatletter
\def\@fpheader{}
\makeatother

\usepackage{graphicx,wrapfig,float,slashed}
\usepackage{amsmath,amssymb,dsfont,epsfig,graphicx,xcolor}
\usepackage{mathtools} 
\usepackage{multirow}
\usepackage{blkarray}
\usepackage{epstopdf}
\usepackage{braket}
\usepackage{ragged2e}
\usepackage[utf8]{inputenc} 
\usepackage{cancel}
\usepackage[table]{xcolor} 
\usepackage{makecell} 
\usepackage{array} 
\usepackage{colortbl} 
\usepackage[toc]{appendix}
\usepackage[normalem]{ulem}
\usepackage{enumitem}
\usepackage[font=small,skip=1pt]{caption}
\allowdisplaybreaks
\usepackage{pifont}
\usepackage{subcaption}
\usepackage{footnote}

\newcommand{\nc}{\newcommand}
\nc{\non}{\nonumber}
\nc{\hc}{\hbox {h.c.}}
\nc{\noi}{\noindent}
\nc{\barx}{\bar{x}}
\nc{\pbarn}{\;\hbox {pb}}
\nc{\fbarn}{\;\hbox {fb}}

\nc{\hsp}{\hspace{0.5cm}}
\nc{\lsp}{\hspace{1cm}}
\nc{\Lsp}{\hspace{2cm}}
\nc{\LLsp}{\lsp\lsp}
\nc{\lra}{\longrightarrow}
\nc{\p}{\prime}
\nc{\sgn}{\text{sgn}}
\nc{\tr}{\text{Tr}}
\nc{\ph}{\varphi}
\nc{\op}{{\cal O}}
\nc{\cL}{{\cal L}}
\nc{\cU}{{\mathcal U}}
\nc{\cD}{{\mathcal D}}
\nc{\cQ}{{\mathcal Q}}
\nc{\cT}{{\mathcal T}}
\nc{\what}{\widehat}
\nc{\vmol}{v_{\text{M\o l}}}

\nc{\beq}{\begin{equation}}  \nc{\eeq}{\end{equation}}
\nc{\bea}{\begin{eqnarray}}  \nc{\eea}{\end{eqnarray}}
\nc{\baa}{\begin{array}}     \nc{\eaa}{\end{array}}
\nc{\bit}{\begin{itemize}}   \nc{\eit}{\end{itemize}}
\nc{\ben}{\begin{enumerate}} \nc{\een}{\end{enumerate}}
\nc{\bce}{\begin{center}}    \nc{\ece}{\end{center}}
\nc{\bpm}{\begin{pmatrix}}   \nc{\epm}{\end{pmatrix}}
\nc{\bvt}{\begin{verbatim}}  \nc{\evt}{\end{verbatim}}

\def\lsim{\mathrel{\raise.3ex\hbox{$<$\kern-.75em\lower1ex\hbox{$\sim$}}}}
\def\gsim{\mathrel{\raise.3ex\hbox{$>$\kern-.75em\lower1ex\hbox{$\sim$}}}}

\def\udots{\mathinner{\mkern1mu\raise1pt\vbox{\kern7pt\hbox{.}}\mkern2mu\raise4pt\hbox{.}\mkern2mu\raise7pt\hbox{.}\mkern1mu}}

\def\tev{\;\hbox{TeV}}

\definecolor{agray}{rgb}{0.95, 0.95, 0.99}

\def\fig#1{Fig.~\ref{#1}}
\def\sec#1{Sec.~\ref{#1}}
\def\tab#1{Table~\ref{#1}}

\def\ohs{{\mathcal O}_{\rm HS}}

\begin{document}

\title{\parbox{\linewidth+6pt}{\hspace{-3pt}General form of effective operators from hidden sectors}}
\author[\star]{Aqeel Ahmed\orcid{0000-0002-2907-2433},}
\emailAdd{aqeel.ahmed@mpi-hd.mpg.de}
\author[\dag]{Zackaria Chacko\orcid{0000-0002-7321-8073},}
\emailAdd{zchacko@umd.edu}
\author[\dag]{Ina Flood\orcid{0000-0002-6897-9820},}
\emailAdd{iflood@umd.edu}
\author[\ddag]{Can Kilic\orcid{0000-0002-9293-1712},}
\emailAdd{kilic@physics.utexas.edu}
\author[\star]{Saereh Najjari\orcid{0000-0003-3053-3759}}
\emailAdd{saereh.najjari@mpi-hd.mpg.de}
\affiliation[\star]{Max-Planck-Institut f\"ur Kernphysik,\\ Saupfercheckweg 1, 69117 Heidelberg, Germany}
\affiliation[\dag]{Maryland Center for Fundamental Physics, Department of Physics,\\ University of Maryland, College Park, MD 20742-4111 USA}
\affiliation[\ddag]{Theory Group, Weinberg Institute for Theoretical Physics\\ University of Texas at Austin, Austin, TX 78712, USA}
\abstract{
We perform a model-independent analysis of the dimension-six terms that are generated in the low energy effective theory when a hidden sector that communicates with the Standard Model (SM) through a specific portal operator is integrated out. We work within the Standard Model Effective Field Theory (SMEFT) framework and consider the Higgs, neutrino and hypercharge portals. We find that, for each portal, the forms of the leading dimension-six terms in the low-energy effective theory are fixed and independent of the dynamics in the hidden sector. For the Higgs portal, we find that two independent dimension-six terms are generated, one of which has a sign that, under certain conditions, is fixed by the requirement that the dynamics in the hidden sector be causal and unitary. 
In the case of the neutrino portal, for a single generation of SM
fermions and assuming that the hidden sector does not violate lepton number, a unique dimension-six term is generated, which corresponds to a specific linear combination of operators in the Warsaw basis. For the hypercharge portal, a unique dimension-six term is generated, which again corresponds to a specific linear combination of operators in the Warsaw basis.
For both the neutrino and hypercharge portals, under certain conditions, the signs of these terms are fixed by the requirement that the hidden sector be causal and unitary. We perform a global fit of these dimension-six terms to electroweak precision observables, Higgs measurements and diboson production data and determine the current bounds on their coefficients.}


\preprint{UTWI-32-2024}
\arxivnumber{2412.15067}

\flushbottom
\maketitle
\flushbottom

\section{Introduction \label{s.introduction}}

While the SM of particle physics provides an excellent description of nature up to energies of order the electroweak scale, it leaves several deep questions unanswered. These include the nature of dark matter, the origin of neutrino masses, the source of the baryon asymmetry, and the resolution of the hierarchy problem. Solutions to these puzzles require new physics beyond the SM.

Experimental data places severe constraints on models of new physics that contain new states at or below the electroweak scale. In general, the types of new physics that are least constrained are those that involve a hidden sector in which none of the new particles carry charges under the SM gauge groups. Such hidden sectors interact with the SM through terms of the form ${\mathcal L}\supset{\mathcal O}_{\rm SM}\ohs$, which couple gauge-singlet SM operators ${\mathcal O}_{\rm SM}$ to operators $\ohs$ in the hidden sector. The corresponding SM operators are referred to as ``portals''. The hidden sector may contain just a small number of weakly coupled states, but it could also contain a multitude of states that have strong interactions with each other and exhibit rich dynamics. Hidden sectors can provide solutions to many of the puzzles of the SM, including the little hierarchy problem, for example,~\cite{Chacko:2005pe,Barbieri:2005ri,Craig:2015pha,Cohen:2018mgv,Cheng:2018gvu}, the origin of the neutrino masses~\cite{Minkowski:1977sc,Yanagida:1979,Gell-Mann:1979vob,Glashow:1979,Mohapatra:1979ia}, the nature of dark matter~\cite{Silveira:1985rk,McDonald:1993ex,Burgess:2000yq,Dodelson:1993je,Pospelov:2007mp,Feng:2008mu} and the origin of the baryon asymmetry~\cite{Fukugita:1986hr,Luty:1992un,Farina:2016ndq,Feng:2020urb,Kilic:2021zqu}.

The three lowest dimensional portal operators in the SM are the Higgs portal ${\mathcal O}_{\rm SM} \equiv H^{\dag}H$, the neutrino portal ${\mathcal O}_{\rm SM} \equiv \ell H$, and the hypercharge portal ${\mathcal O}_{\rm SM} \equiv B_{\mu\nu}$. Therefore, these are the portals most likely to give rise to observable effects. Dedicated searches have been performed at colliders and beam dumps for hidden sector particles that couple through these portals. These include searches for scalar particles that mix with the Higgs boson, heavy neutral leptons that couple through the neutrino portal, and dark photons that mix with the hypercharge gauge boson. 

Even if the states in the hidden sector are too heavy to be directly produced, precision measurements can still provide sensitivity. When integrated out, the heavy hidden sector fields induce higher-dimensional operators that correct the SM Lagrangian. Electroweak precision observables (EWPO) are sensitive to these higher-dimensional operators. The resulting effective operators have been worked out for portals to weakly interacting hidden sectors with a small number of degrees of freedom~\cite{deBlas:2017xtg} and the limits from precision electroweak observables determined, see e.g.~\cite{Ellis:2020unq,Dawson:2020oco,Stefanek:2024kds}. However, the model-dependent approach quickly reaches its limitations when considering more general hidden sectors.

In this paper, we consider a general hidden sector that interacts through the Higgs, neutrino or hypercharge portals. We assume that the portal coupling is small enough to be treated perturbatively and that the states in the hidden sector have masses above the weak scale. We show that, for each of these portals, the forms of the leading dimension-six terms that are generated in the effective Lagrangian when the hidden sector is integrated out are fixed and independent of the dynamics in the hidden sector, only depending on the portal. In particular, for the Higgs portal, we find that two independent dimension-six terms are generated. In the standard Warsaw basis~\cite{Grzadkowski:2010es} for the SMEFT, these take the form, 
 \begin{align}
\mathcal{O}_H=(H^\dag H)^3 \; \; \; \; 
{\rm and} \; \; \; \; 
\mathcal{O}_{H\Box}=(H^\dag H)\Box(H^\dag H) \; .
 \end{align}
 For some range of scaling dimensions of the operator $\ohs$, the sign of the coefficient of the $\mathcal{O}_{H\Box}=(H^\dag H)\Box(H^\dag H)$ term is fixed by the requirement that the dynamics in the hidden sector be causal and unitary. There is no such restriction on the sign of the coefficient of the $\mathcal{O}_H=(H^\dag H)^3$ operator. In the case of the neutrino portal, for a single generation of SM fermions and assuming that the hidden sector does not violate lepton number, a unique dimension-six operator is generated,
  \begin{align}
\mathcal{O}_{\ell H}=({\ell}H)^{\dagger} i\bar{\sigma}^{\mu} \partial_{\mu}( \ell H) \; ,
  \end{align}
where we employ the 2-component spinor conventions described in Ref.~\cite{Dreiner:2008tw}. This operator corresponds to a specific linear combination of two independent operators in the Warsaw basis. For some range of scaling dimensions of the operator $\ohs$, the coefficient of $\mathcal{O}_{\ell H}$ has a definite sign that is fixed by causality and unitarity. 
For the case of the hypercharge portal, again a unique dimension-six term is generated,
 \begin{align}
 \mathcal{O}_{2B}=-\frac{1}{2}(\partial_\rho B_{\mu\nu})(\partial^\rho B^{\mu\nu}) \;.
 \end{align}
This operator, which generates a contribution to the $Y$ parameter~\cite{Barbieri:2004qk} in universal theories~\cite{Wells:2015uba}, corresponds to a specific linear combination of several different operators in the Warsaw basis\,\footnote{Note that our definition of the operator $\mathcal{O}_{2B}$ differs from the commonly used SILH basis operator $\mathcal{O}_{2B}^{\rm SILH}=-(\partial^\mu B_{\mu\nu})(\partial_\rho B^{\rho\nu})/2$~\cite{Giudice:2007fh}. However, after integrating by parts, the two can be seen to be related to each other as $\mathcal{O}_{2B}=2\mathcal{O}_{2B}^{\rm SILH}$.}. For some range of scaling dimensions of the operator $\ohs$, considerations of causality and unitarity again fix the sign of its coefficient.

It is important to note that the restrictions we obtain on the signs of the coefficients of the operators $\mathcal{O}_{H\Box}$, $\mathcal{O}_{\ell H}$ and $\mathcal{O}_{2B}$ are only applicable to interactions of the specific forms we are considering and are not expected to be valid for more general ultraviolet completions. In particular, it is known that the coefficient of the operator $\mathcal{O}_{H\Box}$ can have either sign, depending on the field content of the ultraviolet theory~\cite{Low:2009di}. Several authors have used unitarity and causality to place restrictions on the signs of operators in the SMEFT that are completely general and valid for any ultraviolet completion, for example~\cite{Remmen:2019cyz,Zhang:2020jyn}, (for a review with many additional references see~\cite{deRham:2022hpx}). However, this is usually only possible for dimension-eight operators, which are expected to be less important for most processes at low energies. Therefore, rather than aim for complete generality, we choose to focus on the effective operators at dimension six that arise from a specific class of well-motivated ultraviolet completions. Our philosophy in this regard is similar to that of Ref.~\cite{McCullough:2023szd}, which considered extensions of the SM based on universal theories and showed that the signs of the $Y$ and $Z$ parameters are positive. 

For each portal, we perform a global fit of the dimension-six terms to EWPO, Higgs measurements and diboson data and determine the current bounds on their coefficients. In doing so, we take into account the effects of renormalization group evolution from the matching scale down to the weak scale. The fits find no significant preference for a hidden sector coupled through the Higgs, neutrino, or hypercharge portals over the SM. 

The outline of this paper is as follows. In Section~\ref{s.theory}, we determine the forms of the leading dimension-six operators that arise when a hidden sector that couples to the SM through a specific portal operator is integrated out, and we study the implications of causality and unitarity for the signs of their coefficients. In Section~\ref{s.constraints} we determine the current bounds on these operators from EWPO, Higgs measurements, and diboson production data. We conclude in Section~\ref{s.summary}.

\section{Higher Dimensional Operators from Hidden Sectors \label{s.theory}}

In this section, we determine the leading dimension-six operators that arise in the low-energy effective field theory when a hidden sector that couples to the SM through the Higgs, neutrino or hypercharge portals is integrated out.  We then express these operators in the Warsaw basis and consider their implications for phenomenology.

\subsection{The Higgs Portal \label{s.th.higgs}}

The Higgs portal operator $H^{\dagger}H$ is the lowest dimension gauge invariant scalar operator composed of the SM fields. Interactions of the SM with a hidden sector through the Higgs portal can be written as, 
\begin{equation}
    {\mathcal L}\supset -\lambda H^{\dag}H\mathcal{O}_S,
\label{HPortal1}
\end{equation}
 where $\mathcal{O}_S$ is a gauge invariant scalar operator in the hidden sector. We make no further assumptions about the nature of the operator $\mathcal{O}_S$. It could, for example, represent an elementary or composite operator of a weakly coupled hidden sector, or a primary or secondary operator of a strongly coupled theory that is conformal in the ultraviolet. The coupling $\lambda$ is assumed to be small enough to be treated perturbatively. If the states in the hidden sector have masses above the weak scale, they may be integrated out, giving rise to corrections to the SM Lagrangian. We wish to determine the form of the resulting terms in the low energy effective theory without making any assumptions about the dynamics in the hidden sector. 

At order $\lambda$, the leading effect is a correction to the Higgs propagator. Since there is no momentum flow through the vertex, the only effect that arises is a renormalization of the Higgs mass term, $\mu^2 H^{\dag}H$. However, this has no observable consequences, since the value of $\mu^2$ is in any case not known a priori.  

\begin{figure}
	\centering
        \begin{subfigure}{.5\textwidth}
        \centering
	\includegraphics[width=.6\linewidth]{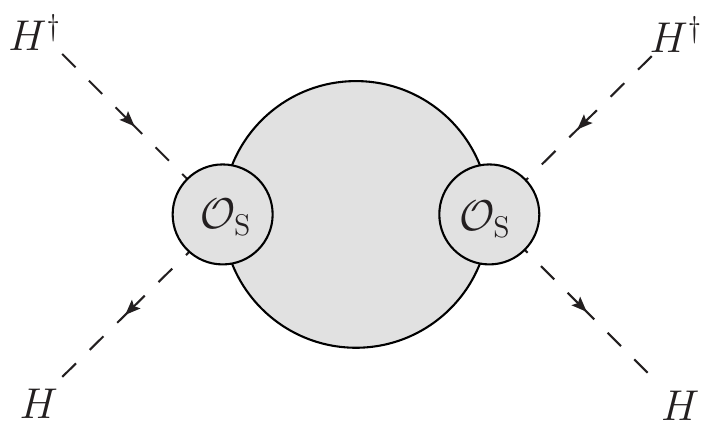}
        \end{subfigure}%
        \begin{subfigure}{.5\textwidth}
        \centering
        \includegraphics[width=.6\linewidth]{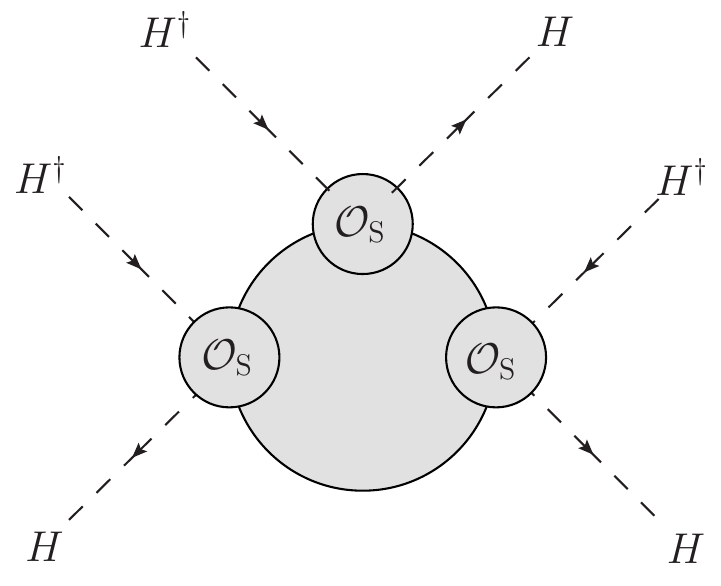}
	\end{subfigure}%
        \caption{Schematic diagrams representing the integrating out of a hidden sector coupled through the Higgs portal at order $\lambda^2$ (left) and $\lambda^3$ (right).}
	\label{fig:HP_Hportal}
\end{figure}

We therefore turn our attention to the corrections generated at order $\lambda^2$. For the purposes of matching, we consider the process $H(p_1),H(p_2) \rightarrow H(p_3),H(p_4)$, shown schematically in Fig.~\ref{fig:HP_Hportal}. The corresponding matrix element takes the form
\begin{align}
   & \frac{1}{2}\bra{p_3,p_4}T\{(-i\lambda)^2\int{d^4x}\int{d^4y[H^{\dag}H\mathcal{O}_{S}](x)[H^{\dag}H\mathcal{O}_{S}](y)}\}\ket{p_1,p_2} \nonumber\\ 
    &= -\frac{\lambda^2}{2}\int{d^4x}\int{d^4y\bra{\Omega}T\{\mathcal{O}_{S}(x)\mathcal{O}_{S}(y) \} \ket{\Omega} \bra{p_3,p_4}T\{H^{\dag}H(x)H^{\dag}H(y)}\}\ket{p_1,p_2},
\label{eq:HP4ptfunc}
\end{align}
where the $p_i$ refer to the external momenta on the Higgs lines, and $\ket{\Omega}$ is the vacuum state.

The hidden sector dynamics is contained in the matrix element $\bra{\Omega}T\{\mathcal{O}_{S}(x)\mathcal{O}_{S}(y) \} \ket{\Omega}$. To simplify this term, we insert a complete set of hidden sector energy-momentum eigenstates $\ket{n}$ with four-momenta $p_n$, 
\begin{align}
    \bra{\Omega}T\{\mathcal{O_S}(x)\mathcal{O_S}(y) \} \ket{\Omega} &= \sum_n T\{ \bra{\Omega}\mathcal{O_S}(x)\ket{n}\bra{n}\mathcal{O_S}(y)\ket{\Omega}\} \nonumber \\
    &= \int{\frac{d^4k}{(2\pi)^3} \rho(k)(\theta(x^0-y^0)e^{-ik\cdot(x-y)}+\theta(y^0-x^0)e^{-ik\cdot(y-x)})}.
\end{align}
In deriving the second line we have used the translation invariance of the vacuum state $\ket{\Omega}$ to set $\bra{\Omega}\mathcal{O_S}(x)\ket{n} = \bra{\Omega}\mathcal{O_S}(0)\ket{n}e^{-ip_n\cdot x} $ and defined the spectral density function
\begin{equation}
    \rho(k)=\sum_n(2\pi)^3\delta^4(p_n-k)\left|\bra{\Omega}\mathcal{O}_{S}(0)\ket{n}\right|^2.
\end{equation}
Note that, by construction, $\rho(k)$ is real and positive. This can be understood as a consequence of unitarity~\cite{Zwicky:2016lka}. 
We now show that $\rho(k)$ transforms as a scalar under Lorentz transformations. Consider $\rho(k')$, where $k'$ is related to $k$ by a Lorentz transformation, $k'^\mu = {\Lambda^\mu}_\nu k^\nu$. Then,
\begin{equation}
    \rho(k')=
    \sum_m(2\pi)^3\delta^4(p_m- \Lambda k)\left|\bra{\Omega}\mathcal{O}_{S}(0)\ket{m}\right|^2.
\end{equation}
The states $\ket{m}$ are related to the states $\ket{n}$ by an (active) unitary transformation, $\ket{m}=\ket{U(\Lambda)n}$. Since the delta function is invariant under Lorentz transformations, we can write
\begin{equation}
    \rho(k') = 
    \sum_n(2\pi)^3\delta^4(p_n - k)\left|\bra{\Omega} U (\Lambda) U^{-1}(\Lambda)\mathcal{O}_{S}(0) U(\Lambda) \ket{n}\right|^2.
\end{equation}
Under a Lorentz transformation the vacuum state $\ket{\Omega}$ and the operator $\mathcal{O_S}(0)$ are invariant. It follows that
$\rho(k) = \rho(k')$, and so $\rho(k)$ is invariant under Lorentz transformations. Then, taking into account that the sum is only over physical states, which necessarily have $k^0 \geq 0$, we have that
\begin{equation}
    \rho(k) = \rho(k^2)\theta(k_0)\,.
\end{equation}
From this we obtain the Kallen-Lehmann spectral representation, 
\begin{align}
  &\bra{\Omega}T\{\mathcal{O_S}(x)\mathcal{O_S}(y) \} \ket{\Omega} \nonumber\\ &=  \int{\frac{d^4k}{(2\pi)^3} (\theta(x^0-y^0)e^{-ik\cdot(x-y)}+\theta(y^0-x^0)e^{-ik\cdot(y-x)})\theta(k_0)\rho(k^2)}\,, \nonumber \\
  &=  \int_0^{\infty}{dM^2\rho(M^2)D_F(x-y,M^2)}\,.
\end{align}
Here $D_F(x-y,M^2)$ is the Feynman (position space) propagator. The Kallen-Lehmann spectral representation is a special case of a dispersion relation, and as such its form is dictated by causality. The matrix element in Eq.~(\ref{eq:HP4ptfunc}) then leads to a sum of two terms of the form 
\begin{align}
    & -\lambda^2\int{d^4x}\int{d^4y\, e^{ip\cdot(x-y)}\int_0^{\infty}{dM^2\rho(M^2)D_F(x-y,M^2)}}
\end{align}
with $p \equiv p_2 - p_4$ for the first term and $p \equiv p_1 - p_4$ for the second. The position integrals over $x$ and $y$ correspond to taking the Fourier transform of the propagator. Then, omitting an overall energy-momentum conserving delta function, we are left with a sum of two terms of the form
 \begin{equation}
    -\lambda^2\int_0^{\infty}{dM^2\rho(M^2)\frac{i}{p^2-M^2+i\epsilon}} \;.
    \label{integral}
 \end{equation}
Based on our assumption that all the states in the hidden sector have masses much greater than $p^2$, we can expand these terms out as
 \begin{equation}
    +i\lambda^2\int_0^{\infty}{dM^2 \frac{\rho(M^2)}{M^2}\left[\left(1+\frac{(p_2-p_4)^2}{M^2}+\ldots\right) + \left(1+\frac{(p_1-p_4)^2}{M^2}+\ldots\right)\right]}.
    \label{expanded_sdf}
 \end{equation}
While the constant terms in the square brackets represent a correction to the Higgs quartic coupling in the SM, which can simply be absorbed into its a priori unknown value, the $p^2/M^2$ terms correspond to a higher dimensional operator of the form
\begin{equation}
    \alpha\frac{\partial_{\mu}(H^{\dag}H)\partial^{\mu}(H^{\dag}H)}{M_{\rm IR}^2}
    \label{eq.Hportal.NPoperator}
\end{equation}
in the low energy effective Lagrangian. Here $M_{\rm IR}$ represents the mass scale of the hidden sector particles that are being integrated out. The dimensionless coefficient $\alpha$ is positive. 

This operator can be expressed in the standard basis for higher dimensional operators, the Warsaw basis. With a simple integration by parts, we find that the operator of interest in this case can be written in terms of the Warsaw basis operator $\mathcal{O}_{H\Box}$
\begin{equation}
\label{alphadef}
    -\frac{\alpha}{M_{\rm IR}^2} (H^{\dag}H)\Box(H^{\dag}H) \equiv
    -\frac{\alpha}{M_{\rm IR}^2} \mathcal{O}_{H\Box}
    \qquad {\rm with}\qquad \alpha>0.
\end{equation}
The fact that the sign of $\alpha$ is fixed can be traced to the positivity of the spectral density function, which is a consequence of unitarity, and the form of the Kallen-Lehmann spectral representation, which is dictated by causality. 

 In general, loop diagrams involving the SM fields will give rise to other dimension-six SMEFT operators at order ${\lambda}^2$ at the matching scale. However, these effects are loop suppressed and therefore subleading compared to the contribution to $\mathcal{O}_{H\Box}$.   

In performing the analysis leading up to Eq.~(\ref{eq.Hportal.NPoperator}), we have implicitly assumed that the integral in Eq.~(\ref{integral}) does not diverge in the ultraviolet. However, in general, this assumption may not be valid. Consider, for example, the specific case of an operator $\mathcal{O}_S$ of scaling dimension $\Delta_S$ in the ultraviolet, so that $\rho(M^2)$ scales as $(M^2)^{\Delta_S - 2}$ for large $M^2$. Then the integral in Eq.~(\ref{integral}) is ultraviolet-divergent for $\Delta_S \geq 2$ and must be regulated by adding counterterms. However, as shown in Appendix \ref{appendix1}, for $2 \leq \Delta_S < 3$, a counterterm for the Higgs quartic suffices to regulate the theory to order $\lambda^2$, and the sign of $\alpha$ in Eq.~(\ref{eq.Hportal.NPoperator}) is unaffected. However, for $\Delta_S > 3$, the coefficient $\alpha$ of the higher dimensional term in Eq.~(\ref{eq.Hportal.NPoperator}) is also ultraviolet sensitive at order $\lambda^2$ and receives most of its support from unknown ultraviolet physics. It therefore now requires a counterterm, and our argument that $\alpha > 0$ no longer applies. For the special case of $\Delta_S = 3$, $\alpha$ is only logarithmically divergent, and therefore contributions from scales below the ultraviolet cutoff are logarithmically enhanced. We therefore expect that, although a counterterm is still required to account for the unknown ultraviolet physics, $\alpha > 0$ in this case as well. Hence our conclusion about the sign of $\alpha$ is expected to be valid for the range of scaling dimensions $\Delta_S \leq 3$.      

 This result admits a simple understanding based on dimensional analysis. When the operator $\mathcal{O}_S$ has a definite scaling dimension $\Delta_S$ in the ultraviolet, we can write the Higgs portal interaction in Eq.~(\ref{HPortal1}) as
 \begin{equation}
    {\mathcal L}\supset -\lambda H^{\dag}H\mathcal{O}_S \equiv -\frac{\hat{\lambda}}{M_{\rm UV}^{\Delta_S-2}} H^{\dag}H\mathcal{O}_S \;.
 \end{equation}
 Here $M_{\rm UV}$ represents an ultraviolet scale and the coupling $\hat{\lambda}$ is dimensionless. The higher dimensional term in Eq.~(\ref{eq.Hportal.NPoperator}) is generated at order $\lambda^2$ and its coefficient $\alpha/M_{\rm IR}^2$ has mass dimension $-2$. It then follows that $\alpha/M_{\rm IR}^2$ is finite and scales as $\hat{\lambda}^2(M_{\rm IR}/M_{\rm UV})^{2\Delta_S-4}(1/M_{\rm IR})^2$ for $\Delta_S < 3$, but is ultraviolet-divergent and scales as $\hat{\lambda}^2(\Lambda_{\rm UV}/M_{\rm UV})^{2\Delta_S-4}(1/\Lambda_{\rm UV})^2$, where $\Lambda_{\rm UV}$ is an ultraviolet cutoff, for $\Delta_S \geq 3$. For the special case of $\Delta_S = 3$, the divergence is only logarithmic and $\alpha/M_{\rm IR}^2$ scales as $\hat{\lambda}^2 (1/M_{\rm UV})^2 {\rm ln}(\Lambda_{\rm UV}/M_{\rm IR})$.
 Therefore the sign of $\alpha$ is expected to be positive for $\Delta_S \leq 3$. 
 
 In the more general case, $\mathcal{O}_S$ may not have a definite scaling dimension in the ultraviolet. For example, this would be the case if $\mathcal{O}_S$ consists of a linear combination of operators of definite scaling dimension.  As shown in Appendix \ref{appendix1}, the conclusion that $\alpha$ is positive is then satisfied provided that $\rho(M^2)$ does not grow any faster than $M^2$ in the ultraviolet. 
 
We now turn our attention to the leading dimension-six operators generated at order~$\lambda^3$. The only such operator is $(H^{\dag}H)^3$, which arises from diagrams of the schematic form shown in Fig.~\ref{fig:HP_Hportal}. This operator is represented by $\mathcal{O}_{H}$ in the Warsaw basis. In contrast to $\mathcal{O}_{H\Box}$, there is no restriction on the sign of this operator. While loop diagrams involving the SM fields can generate other dimension-six operators at order $\lambda^3$ at the matching scale, these effects are loop-suppressed and therefore subleading.  

We now consider the leading physical effects of these operators. In the case of $\mathcal{O}_{H\Box}$, going to unitary gauge, we see that the kinetic term of the physical Higgs boson receives a correction of order  $\alpha v_{\rm EW}^2 / M_{\rm IR}^2$. After rescaling the Higgs field to return to the canonical normalization, the couplings of the Higgs field to the SM fermions and gauge bosons all receive a common correction of this order. This alters the production rates of the Higgs boson at the Large Hadron Collider (LHC), but not its branching fractions. The bounds on the Higgs event rates can therefore be used to constrain $\mathcal{O}_{H\Box}$. 

In the case of $\mathcal{O}_H$, its main effect is to correct the potential for the Higgs doublet in the SM. After setting the mass of the physical Higgs boson to the observed value, its trilinear term and quartic couplings receive corrections that cause them to differ from the SM predictions. At present, these couplings are poorly constrained by the data and so the bounds on this operator are rather weak. We discuss the constraints on both these operators in greater detail in Section~\ref{s.constraints}.

\subsection{The Neutrino Portal \label{s.th.neutrino}}

The neutrino portal interaction can be written as, 
 \begin{equation}
    {\mathcal L}\supset -y \mathcal{O}_F \ell H + {\rm h.c.}\;,
\label{neutrinoportal}
 \end{equation}
where the SM electroweak doublet lepton field $\ell$ and the hidden sector operator $\mathcal{O}_F$ transform as left-chiral spinors under the Lorentz group. We adopt the convention in which the hypercharge of the Higgs doublet has the opposite sign to that of the left handed leptons. Then the $SU(2)$ gauge indices on these fields are contracted via the antisymmetric tensor,
$\ell H \equiv \epsilon_{ab} \ell^a H^b$,
where $\epsilon_{ab}= \left(i \sigma^2\right)_{ab}$.
In our analysis, we will assume that the hidden sector does not violate lepton number. For now, we limit our consideration to a single generation of SM fermions.

\begin{figure}
	\centering
	\includegraphics[width=.6\linewidth]{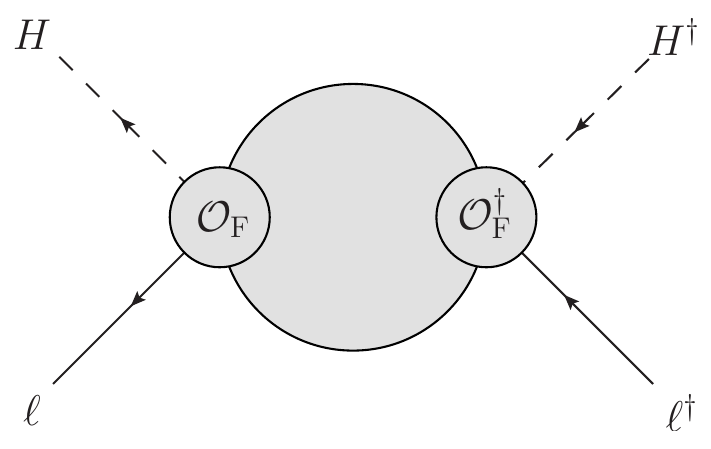}
	\caption{Schematic diagram representing the integrating out of a hidden sector coupled through the neutrino portal at order $y^2$.}
	\label{fig:NP_Nportal}
\end{figure}

We wish to integrate out the hidden sector to determine the form of the dimension-six terms in the low energy effective theory. Since $\mathcal{O}_F$ is not a singlet under Lorentz transformations, the effects arising from integrating out the hidden sector begin at order~$y^2$. For the purposes of matching at this order, we consider the process $\ell(p_1) H(p_2) \rightarrow \ell(p_3) H(p_4)$, represented schematically in Fig.~\ref{fig:NP_Nportal}. The corresponding matrix element is given by
\begin{align}
    \label{lepton_expr}
    & \bra{p_3,x(p_3),p_4}T\left\{(-iy)^2\int{d^4x}\int{d^4y[(\ell H)^{\dag}
    {\mathcal{O}_F}^\dagger](x)[\mathcal{O}_F \ell H](y)}\right\}\ket{p_1,x(p_1),p_2}\nonumber = \\
    &(-iy)^2 \int{d^4x}\int{d^4y\bra{p_3,x(p_3),p_4}T\left\{(\ell H)^{\dag}(x)\ell H(y)\right\}\ket{p_1,x(p_1),p_2}} \bra{\Omega}T\{{\mathcal{O}_F}^\dagger(x)\mathcal{O}_F(y)\}\ket{\Omega},
\end{align}
where $x(p_{1,3})$ are the initial and final state lepton spinors, and we employ the notation of Ref.~\cite{Dreiner:2008tw}.

Once again, we insert a complete set of hidden sector energy-momentum eigenstates $\ket{n}$ inside the matrix element $\bra{\Omega}T\{{\mathcal{O}_F}^\dagger(x)\mathcal{O}_F(y)\}\ket{\Omega}$, and express the result as,
\begin{equation}
    \int{\frac{d^4k}{(2\pi)^3}\left(\theta(x^0-y^0)e^{-ik\cdot(x-y)}-\theta(y^0-x^0)e^{-ik\cdot(y-x)}\right)\rho^{\dot{\alpha}\alpha}(k)}\,,
\label{fermion2pointfunction}    
\end{equation}
where the spectral density function 
\begin{equation}
    \rho^{\dot{\alpha}\alpha}(k) = (2\pi)^3\sum_n\delta^4(p_n-k)\bra{\Omega}\mathcal{O}_F^{\dagger\dot{\alpha}}(0)\ket{n}\bra{n}\mathcal{O}_F^{\alpha}(0)\ket{\Omega} \;.
\end{equation}
The two terms in Eq.~(\ref{fermion2pointfunction}), which each correspond to a different time-ordering, are related to each other by $\mathcal{CPT}$ symmetry, as discussed in Appendix~\ref{appendix2}.
We now determine the behavior of $\rho^{\dot{\alpha}\alpha}(k)$ under a Lorentz transformation.  Consider $\rho^{\dot{\alpha}\alpha}(k')$, where $k'$ is related to $k$ by a Lorentz transformation, $k'^\mu = {\Lambda^\mu}_\nu k^\nu$. Then,
\begin{equation}
    \rho^{\dot{\alpha}\alpha}(k') = \sum_m(2\pi)^3\delta^4(p_m- \Lambda k)
    \bra{\Omega}\mathcal{O}_F^{\dagger\dot{\alpha}}(0)\ket{m}\bra{m}\mathcal{O}_F^{\alpha}(0)\ket{\Omega},
\end{equation}
where the states $\ket{m}$ are related to the states $\ket{n}$ by an active unitary transformation, $\ket{m}=\ket{U(\Lambda)n}$. Since the delta function is invariant under Lorentz transformations, we can write
\begin{eqnarray}
    \rho^{\dot{\alpha}\alpha}(k') &=& \sum_n(2\pi)^3\delta^4(p_n - k) 
       \bra{\Omega}\mathcal{O}_F^{\dagger\dot{\beta}}(0) \ket{Un}\bra{Un}\mathcal{O}_F^{\beta}(0) \ket{\Omega}\,, \nonumber \\
       &=&
       \sum_n(2\pi)^3\delta^4(p_n - k) 
       \bra{\Omega} U U^{-1}\mathcal{O}_F^{\dagger\dot{\beta}}(0) U\ket{n}\bra{n} U^{-1}\mathcal{O}_F^{\beta}(0) U U^{-1} \ket{\Omega}\,.
\end{eqnarray}
Under an active Lorentz transformation the vacuum is invariant, while the fermionic operators transform as
\begin{align}
    U^{-1}(\Lambda)\mathcal{O}_F^{\dagger\dot{\alpha}}(0) U(\Lambda) &=   
    [(\mathcal{M}^{-1}(\Lambda))^{\dagger}]^{\dot{\alpha}}_{\text{ }\dot{\beta}}\mathcal{O}_F^{\dagger\dot{\beta}}(0)\;, \\  
    U^{-1}(\Lambda)\mathcal{O}_F^{\alpha}(0)U(\Lambda) &=
    \mathcal{O}_F^{\beta}(0)[\mathcal{M}^{-1}(\Lambda)]^{\text{ }\alpha}_{\beta}
    \;,
\end{align}
where $\mathcal{M}$ is the representation matrix of the $(\frac{1}{2},0)$ Lorentz algebra. Then
\begin{equation}
        \rho^{\dot{\alpha}\alpha}(k')
       =   [(\mathcal{M}^{-1}(\Lambda))^{\dagger}]^{\dot{\alpha}}_{\text{ }\dot{\beta}}\left[\sum_n(2\pi)^3\delta^4(p_n-k) 
     \bra{\Omega}\mathcal{O}_F^{\dagger\dot{\beta}}(0)\ket{n}\bra{n}\mathcal{O}_F^{\beta}(0)\ket{\Omega}\right] [\mathcal{M}^{-1}(\Lambda)]^{\text{ }\alpha}_{\beta} \;.
\end{equation}
Recognizing the term in square brackets as $\rho^{\dot{\beta}\beta}(k)$, we have that under Lorentz transformations the spectral density function transforms as
\begin{equation}
    \label{lepton_constraint}
    \rho^{\dot{\alpha}\alpha}(\Lambda k)=[(\mathcal{M}^{-1}(\Lambda))^{\dagger}]^{\dot{\alpha}}_{\text{ }\dot{\beta}}\rho^{\dot{\beta}\beta}(k)[\mathcal{M}^{-1}(\Lambda)]^{\text{ }\alpha}_{\beta}\,.
\end{equation}
Since $\rho^{\dot{\alpha}\alpha}$ is a ($2 \times 2$) Hermitian matrix, we can expand it out in the basis of the $\bar{\sigma}^\mu$ matrices. From the spinor structure of $\rho(k)$, we have
\begin{equation}
    \rho^{\dot{\alpha}\alpha}(k)=\rho_{\mu}(k)(\bar{\sigma}^{\mu})^{\dot{\alpha}\alpha}\,.
\end{equation}
Under Lorentz transformations the $\bar{\sigma}$ matrices transform as
\begin{equation}
    (\mathcal{M}^{-1})^{\dagger}\bar{\sigma}^{\mu}\mathcal{M}^{-1} =\Lambda_{\nu}^{\text{ }\mu} \bar{\sigma}^{\nu} \;.
\end{equation}
 Then, from Eq.~(\ref{lepton_constraint}), $\rho_{\mu}(k)$ must transform as a covariant vector under Lorentz transformations. Since $\rho^{\dot{\alpha}\alpha}$ is a function of $k$ alone and is constructed from a sum over physical states, we have that $\rho_\mu(k)$ must be of the form $\rho_{\mu}(k)= k_{\mu} \rho(k^2)\theta(k_0)$. Therefore the matrix element in Eq.~(\ref{lepton_expr}) can be simplified to
\begin{equation}
    (-iy)^2x^{\dag}(p_3)\left[\int_0^\infty{dM^2\frac{\rho(M^2)(ip_{\mu}\bar{\sigma}^{\mu})}{p^2-M^2+i\epsilon}}\right]x(p_1),
    \label{eq:NportalMatrixElement}
\end{equation}
In this expression $p = (p_1 + p_2)$ and $\rho(k^2) \theta(k_0)$ is given by
\begin{equation}
    \rho(k^2) \theta(k_0) = \frac{(2\pi)^3}{2k^0} \sum_n\delta^4(q_n-k)\sum_{\alpha}\left|\bra{n}{\mathcal O}^\alpha_{F}(0)\ket{\Omega}\right|^2 \; .
    \label{eq:NportalRhoDef}
\end{equation}
Expanding in powers of $p^2/M^2$, the leading term in Eq.~(\ref{eq:NportalMatrixElement}) is given by
\begin{align}
    & iy^2\,x^{\dag}(p_3) (p_{\mu}\bar{\sigma}^{\mu}) x(p_1) \int_0^\infty{\frac{dM^2}{M^2}\rho(M^2)} \;.
\end{align}
This corresponds to a dimension-six operator in the low energy effective theory of the form
\begin{equation}
    \frac{\alpha}{M^{2}_{\rm IR}}({\ell} H)^{\dagger} i\bar{\sigma}^{\mu} \partial_{\mu}(\ell H)  
   \equiv \frac{\alpha}{M^{2}_{\rm IR}} \mathcal{O}_{\ell H} \;.
    \label{eq.Nportal.NPoperator}
\end{equation}
The positivity of $\rho(k^2)$ in Eq.~(\ref{eq:NportalRhoDef}) dictates that the dimensionless coefficient $\alpha$ in this equation is necessarily positive. This is again a consequence of the requirement that the hidden sector dynamics be causal and unitary. In general, loop diagrams involving the SM fields will give rise to other dimension-six SMEFT operators at the matching scale at order ${y}^2$. However, these effects are suppressed by a loop factor and therefore subleading compared to the operator in Eq.~(\ref{eq.Nportal.NPoperator}).
 
We have determined that the coefficient of the operator $\mathcal{O}_{\ell H}$ in the low energy effective theory is necessarily positive. However, in performing the steps leading up to Eq.~(\ref{eq.Nportal.NPoperator}), we have implicitly assumed that the integral in Eq.~(\ref{eq:NportalMatrixElement}) does not diverge in the ultraviolet, and this assumption must be reexamined. For example, consider the specific case of an operator $\mathcal{O}_F$ of scaling dimension $\Delta_F$ in the far ultraviolet. Then $\rho(M^2)$ scales as $(M^2)^{\Delta_F - 5/2}$ for large $M^2$. The integral is ultraviolet-divergent for $\Delta_F \geq 5/2$ and must be regulated by adding a counterterm for $\mathcal{O}_{\ell H}$. Then, for $\Delta_F > 5/2$, the coefficient $\alpha$ of the higher dimensional term in Eq.~(\ref{eq.Nportal.NPoperator}) receives most of its support from unknown ultraviolet physics and our argument that $\alpha > 0$ no longer applies. For the special case of $\Delta_F = 5/2$, $\alpha$ is only logarithmically divergent, and therefore contributions from scales below the ultraviolet cutoff are logarithmically enhanced. We therefore expect that $\alpha > 0$ in this case as well. Hence our conclusion that $\alpha$ is positive is expected to be valid for the range of scaling dimensions $\Delta_F \leq 5/2$. In the more general case, $\mathcal{O}_{F}$ may not have a definite scaling dimension in the ultraviolet. In such a case, the conclusion that $\alpha$ is positive is satisfied provided that $\rho(M^2)$ does not increase with $M^2$ in the ultraviolet, but instead falls or remains constant.  

We express the operator in $\mathcal{O}_{\ell H}$ in the Warsaw basis, for later use in Section~\ref{s.constraints}. After some algebra, we arrive at the linear combination
\begin{equation}
    \mathcal{O}_{\ell H}=\frac14\left[ \mathcal{O}_{H \ell }^{(1)}- \mathcal{O}_{H \ell }^{(3)}\right] \;,      \label{eq.NPWarsawbasis}
\end{equation}
where the Warsaw basis operators $\mathcal{O}_{H \ell }^{(1)}$ and $\mathcal{O}_{H \ell }^{(3)}$ (omitting generation indices) are defined as
\begin{eqnarray}
    \mathcal{O}_{H \ell }^{(1)}& \equiv&(H^{\dagger} i \stackrel{\leftrightarrow}{D}_{\mu} H)(\bar{\ell} \gamma^{\mu} \ell),\nonumber\\
    \mathcal{O}_{H \ell }^{(3)} &\equiv&(H^{\dagger} i \stackrel{\leftrightarrow}{D}_{\mu}^{I} H)(\bar{\ell} \tau^{I} \gamma^{\mu} \ell).
\end{eqnarray}

We now consider the realistic case of three generations of SM fermions. In general, there could also be multiple flavors of fermionic operators $\mathcal{O}_{F, \alpha}$ in the hidden sector, where $\alpha$ is a flavor index for the hidden sector operators. Then Eq.~(\ref{neutrinoportal}) generalizes to
 \begin{equation}
    {\mathcal L}\supset -y^{\alpha i} \mathcal{O}_{F, \alpha} \ell_i H \equiv -\frac{\hat{y}^{\alpha i}}{M_{\rm UV}^{\Delta_F-3/2}} \mathcal{O}_{F, \alpha} \ell_i H \;,
\label{neutrinoportal3}
 \end{equation}
where $i = 1,2,3$ is a SM flavor index. After integrating out the hidden sector states, we obtain
\begin{equation}
    \frac{\alpha^{ij}}{M^{2}_{\rm IR}}({\ell}_i H)^{\dagger} i\bar{\sigma}^{\mu} \partial_{\mu}(\ell_j H) \;.
    \label{eq.Nportal.NPoperator3}
\end{equation}
This is the generalization of Eq.~(\ref{eq.Nportal.NPoperator}) to the three flavor case. If the operators $\mathcal{O}_{F, \alpha}$ all have the same scaling dimension $\Delta_F$, and are also orthogonal in flavor space, so that 
$\langle\mathcal{O}_{F, \alpha}(x)\mathcal{O}^\dagger_{F, \beta}(y) \rangle \propto \delta_{\alpha \beta}$, then the matrix composed of the $\alpha^{ij}$ in Eq.~(\ref{neutrinoportal3}) is expected to be positive definite for $\Delta_F \leq 5/2$.

To understand the leading phenomenological effects of the operator in Eq.~(\ref{eq.Nportal.NPoperator3}), we set the Higgs to its VEV. This results in corrections to the kinetic terms of the neutrinos that scale as $\alpha v_{\rm EW}^2 / M_{\rm IR}^2$, without any corresponding change in the kinetic terms of the charged leptons. After rescaling the kinetic terms of the neutrinos to their canonical values, this will result in corrections to the couplings of the neutrinos to the $W$ and $Z$ bosons. Such corrections are very strongly constrained because of the high precision in the measured values of observables such as the muon lifetime and the $Z$-boson line shape. We will determine the constraints on the effective operator in Eq.~(\ref{eq.Nportal.NPoperator3}) in Section~\ref{s.constraints}, under the assumption that the couplings to the 
hidden sector are flavor universal, so that $\alpha^{ij} \propto \delta^{ij}$.

If the $\alpha^{ij}$ are not flavor diagonal, the couplings of the neutrinos to the weak gauge bosons violate the lepton flavor symmetries of the SM. This will give rise to charged lepton flavor violating processes such $\mu \rightarrow e \gamma$ and $\mu \rightarrow 3 e$, which are very tightly constrained by data. We leave a careful study of these effects for future work.

\subsection{The Hypercharge Portal \label{s.th.hypercharge}}

\begin{figure}
	\centering
	\includegraphics[width=.6\linewidth]{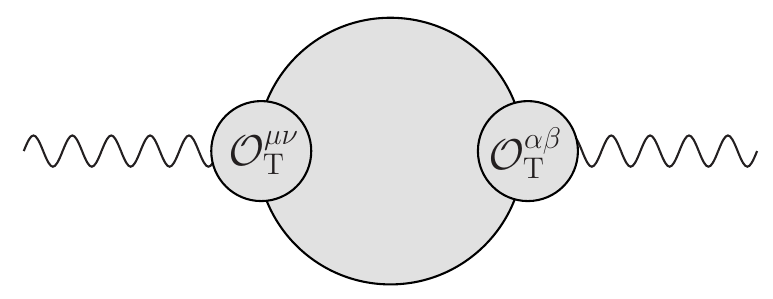}
	\caption{Schematic diagram representing the integrating out of a hidden sector coupled through the hypercharge portal at order $\epsilon^2$.}
	\label{fig:NP_Bportal}
\end{figure}

Interactions of the SM with a hidden sector through the hypercharge portal take the form,
\begin{equation}
    \mathcal{L} \supset - \epsilon B_{\mu\nu}\mathcal{O}_T^{\mu\nu} \;.
\end{equation}
Here $\mathcal{O}_T^{\mu\nu}$ is a real antisymmetric rank-2 tensor operator in the hidden sector. As before, we wish to determine the forms of the dimension-six operators that are generated when the hidden sector is integrated out. As in the case of the neutrino portal, since $\mathcal{O}_T^{\mu\nu}$ is not a singlet under Lorentz transformations, these effects can only arise at order $\epsilon^2$ or higher. The leading effects then arise from corrections to the propagator of the hypercharge gauge boson, shown schematically in Fig.~\ref{fig:NP_Bportal}. 

At order $\epsilon^2$, the correction to the $B_{\mu}$ 2-point function takes the form
\begin{eqnarray}
   &&\frac{1}{2}\bra{0}T\Bigl\{B_\rho(x')B_\sigma(y')  (-i\epsilon)^2\int{d^4x}\int{d^4y} [B_{\mu\nu}\mathcal{O}_T^{\mu\nu}](x)[B_{\alpha\beta}\mathcal{O}_T^{\alpha\beta}](y)\Bigr\}\hspace{0.5em}\ket{0} = \\
     &&  -\frac{\epsilon^2}{2} 
    \int{d^4x}\int{d^4y}\bra{\Omega}T\left\{\mathcal{O}_T^{\mu\nu}(x)\mathcal{O}_T^{\alpha\beta}(y)\right\}\ket{\Omega} 
    \bra{0}T\left\{B_\rho(x')B_\sigma(y')B_{\mu\nu}(x)B_{\alpha\beta}(y)\right\} \ket{0}. \nonumber
    \label{eq.Bportal.2ptfunc}
\end{eqnarray}
The matrix element involving the gauge boson $B_{\mu}$ and the field strength $B_{\mu\nu}$ is straightforward to evaluate. We therefore start by focusing on the hidden sector matrix element. As before, we insert a complete set of energy-momentum eigenstates and use translation invariance to separate out the position dependence,
\begin{align}
     &\bra{\Omega}T\left\{\mathcal{O}_T^{\mu\nu}(x)
     \mathcal{O}_T^{\alpha\beta}(y)\right\}\ket{\Omega}  \nonumber\\
    =& \sum_{n} \theta(x^0-y^0)e^{-ip_n.(x-y)}\bra{\Omega}
    \mathcal{O}_T^{\mu\nu}(0)\ket{n}\bra{n}\mathcal{O}_T^{\alpha\beta}(0)\ket{\Omega} 
    \notag\\
    & +\sum_{n}
    \theta(y^0-x^0)e^{-ip_n.(y-x)}\bra{\Omega}
    \mathcal{O}_T^{\alpha\beta}(0)\ket{n}\bra{n}\mathcal{O}_T^{\mu\nu}(0)\ket{\Omega}\,,
    \nonumber\\
    =& \int \frac{d^4q}{(2\pi)^3}\left\{\theta(x^0-y^0)e^{-iq.(x-y)}\pi^{\mu\nu\alpha\beta}(q)
    +\theta(y^0-x^0)e^{-iq.(y-x)}\pi^{\alpha\beta\mu\nu}(q) \right\} \;,
\end{align}
where the tensor $\pi^{\mu\nu\alpha\beta}$ is given by
\begin{equation}
    \pi^{\mu\nu\alpha\beta}(q) \equiv (2\pi)^3 \sum_n \delta^4(p_n-q) \bra{\Omega}
    \mathcal{O}_T^{\mu\nu}(0)\ket{n}\bra{n}
    \mathcal{O}_T^{\alpha\beta}(0)\ket{\Omega} \;.
    \label{eq.pi4tensor.def}
\end{equation}
Because $\mathcal{O}_T$ is a real, antisymmetric tensor, $\pi^{\mu\nu\alpha\beta}$ is antisymmetric under the exchange of the indices, $\mu\longleftrightarrow\nu$ and $\alpha\longleftrightarrow\beta$. Then Lorentz symmetry constrains  $\pi^{\mu\nu\alpha\beta}(q)$ to be of the general form
\begin{equation}
\pi^{\mu\nu\alpha\beta}(q)=\rho_{\epsilon}(q^2)\epsilon^{\mu\nu\alpha\beta}+\rho_{0}(q^2)\Pi_{0}^{\mu\nu\alpha\beta}+\rho_{1}(q^2)\Pi_{1}^{\mu\nu\alpha\beta}(q),
\label{eq.pi4tensor.rhos}
\end{equation}
where $\epsilon^{\mu\nu\alpha\beta}$ is the Levi-Civita tensor, and
\begin{eqnarray}
    \Pi_{0}^{\mu\nu\alpha\beta}&\equiv& g^{\mu\alpha}g^{\nu\beta}-g^{\mu\beta}g^{\nu\alpha},\nonumber\\
    \Pi_{1}^{\mu\nu\alpha\beta}(q)&\equiv& - g^{\mu\alpha}q^\nu q^\beta +g^{\mu\beta}q^\nu q^\alpha -g^{\nu\beta}q^\mu q^\alpha +g^{\nu\alpha}q^\mu q^\beta \;.
    \label{eq.pi4tensor.Pis}
\end{eqnarray}
It follows from this general structure that the tensor $\pi^{\mu\nu\alpha\beta}$ is symmetric under the exchange of pairs of indices, $(\mu\nu)\longleftrightarrow(\alpha\beta)$, so $\pi^{\mu\nu\alpha\beta}(q) = \pi^{\alpha\beta\mu\nu}(q)$. This relation can also be understood as arising from $\mathcal{CPT}$ invariance of the theory, as discussed in Appendix~\ref{appendix2}.
This allows us to write the 2-point function in Eq.~(\ref{eq.Bportal.2ptfunc}) as  
\begin{equation}
    \bra{\Omega}T\left\{\mathcal{O}_T^{\mu\nu}(x)
     \mathcal{O}_T^{\alpha\beta}(y)\right\}\ket{\Omega} = \int{dM^2} D_F(x-y,M^2)\int{\frac{d^4q}{(2\pi)^4}}\delta(q^2-M^2) \pi^{\mu \nu \alpha \beta} (q) \; .
\end{equation}

 We now show that the signs of $\rho_{0}$ and $\rho_{1}$ in Eq.~(\ref{eq.pi4tensor.rhos}) are both positive. Consider elements of the tensor $\pi^{\mu\nu\alpha\beta}$ of the form $\pi^{\mu\nu\mu\nu}$ (no sum). From Eq.~(\ref{eq.pi4tensor.def}), each such element is of positive sign. Now consider the case when $q^{\mu}$ is timelike, so that we can choose a frame in which $q^{\mu}$ is along the time direction, $q^{\mu}=\sqrt{q^2}\delta^{\mu}_{0}$. Then, from Eqs.~(\ref{eq.pi4tensor.rhos}) and (\ref{eq.pi4tensor.Pis}), choosing $\mu$ and $\nu$ to be distinct spatial indices $i$ and $j$, $i \neq j$, (so that both $\epsilon^{ijij}$ and $\Pi_{1}^{ijij}$ vanish, while $\Pi_{0}^{ijij}=1$), it is straightforward to see that $\rho_{0}(q^2)$ has positive sign for timelike $q^{\mu}$. Next, choosing $\mu=0$ and $\nu$ to be any spacelike index, we find
\begin{equation}
    \pi^{0i0i} = q^2 \rho_{1}(q^2) - \rho_{0}(q^2)>0 \;.
    \label{eq.pi4tensor.positivity}
\end{equation}
This shows that $\rho_{1}(q^2)$ has positive sign for timelike $q^\mu$.

We are now in a position to contract the hidden sector matrix element with the SM matrix element. Note that the $\rho_{\epsilon}$ term does not contribute, since there are no four linearly independent vectors to contract it with. Working in Feynman gauge, we obtain,
\begin{eqnarray}
    -\epsilon^2 \int_0^\infty dM^2 && \frac{i}{p^2 - M^2 + i\epsilon}\left(\rho_{0}(M^2)\Pi_{0}^{\mu\nu\alpha\beta} + \rho_{1}(M^2)\Pi_{1}^{\mu\nu\alpha\beta}(p)\right) \nonumber \\
    &&  \times \left( p_\mu \frac{g_{\rho\nu}}{p^2+i\epsilon} -p_\nu \frac{g_{\rho\mu}}{p^2+i\epsilon} \right)\left( p_\alpha \frac{g_{\beta\sigma}}{p^2+i\epsilon} -p_\beta \frac{g_{\alpha\sigma}}{p^2+i\epsilon} \right)\nonumber\\
    = 4\epsilon^2 &&\left(g_{\rho\sigma}-\frac{p_\rho p_\sigma}{p^2}\right)\frac{i}{p^2+i\epsilon} \int_0^\infty dM^2 \; \frac{-\rho_{0}(M^2)+p^2\rho_{1}(M^2)}{p^2 - M^2 + i\epsilon} \;.
\end{eqnarray}
where $p$ is the momentum flowing through the propagator, Expanding the integrand out in powers of $p^2/M^2$, we obtain
\begin{align}
    4\epsilon^2 \left(g_{\rho\sigma}-\frac{p_\rho p_\sigma}{p^2}\right)\frac{i}{p^2+i\epsilon}
    \left\{ \int_0^\infty \frac{dM^2}{M^2} \rho_{0}(M^2) - p^2 
    \int_0^\infty \frac{dM^2}{M^4} \left(-\rho_{0}(M^2)+M^2\rho_{1}(M^2)\right) \right\}.
\label{TPortalintegrals}
\end{align}
Note that we get the correct tensor structure, as dictated by gauge invariance. Also, note that both of the integrands in the curly brackets are positive, the first one due to the positivity of $\rho_0(q^2)$ and the second one from the positivity of the expression in Eq.~(\ref{eq.pi4tensor.positivity}). The first integral in the curly brackets represents a contribution to the kinetic term of the hypercharge gauge boson. It has no observable consequences since it can be absorbed into the a priori unknown value of the $U(1)_Y$ gauge coupling. The second term, on the other hand, has observable effects and corresponds in the low energy effective theory to a dimension-six operator of the form
\begin{equation}
\frac{\alpha}{M_{\rm IR}^2}\mathcal{O}_{2B}\qquad {\rm where}\qquad \alpha>0\qquad {\rm and}\qquad \mathcal{O}_{2B}\equiv -\frac{1}{2}(\partial_{\lambda}B_{\mu\nu})(\partial^{\lambda}B^{\mu\nu}).
    \label{eq.Bportal.NPoperator}
\end{equation}
 Here $\alpha$ is again dimensionless.

It is the positivity of the integrand in the second integral in Eq.~(\ref{TPortalintegrals}) that led us to conclude that $\alpha$ is positive. However, in reaching this conclusion we have implicitly assumed that the integrals in Eq.~(\ref{TPortalintegrals}) do not diverge in the ultraviolet, and this assumption must be reexamined. Consider an operator $\mathcal{O}_T^{\mu \nu}$ of scaling dimension $\Delta_T$ in the ultraviolet. Then $\rho_0(M^2)$ and $\rho_1(M^2)$ scale as $(M^2)^{\Delta_T - 2}$ and $(M^2)^{\Delta_T - 3}$ for large $M^2$. Hence the first integral in Eq.~(\ref{TPortalintegrals}) is ultraviolet-divergent for $\Delta_T \geq 2$ and must be regulated. This can be done by adding a counterterm for the kinetic term for the hypercharge gauge boson. For $\Delta_T \geq 3$, the second integral in Eq.~(\ref{TPortalintegrals}) is also ultraviolet-divergent. For $\Delta_T > 3$ the coefficient $\alpha$ of the higher dimensional term in Eq.~(\ref{eq.Nportal.NPoperator}) receives most of its support from the unknown physics at scales of order the cutoff and our argument that $\alpha > 0$ no longer applies. For the special case of $\Delta_T = 3$, $\alpha$ is only logarithmically divergent, and therefore contributions from scales below the ultraviolet cutoff are logarithmically enhanced. We therefore expect our argument that $\alpha > 0$ to apply to this case as well. Hence our conclusion that $\alpha$ is positive is expected to be valid for the range of scaling dimensions $\Delta_T \leq 3$. In the more general case, the operator $\mathcal{O}_{T}$ may not have a definite scaling dimension in the ultraviolet. The conclusion that $\alpha$ is positive is then satisfied provided that the combination $\rho_0(M^2) - M^2 \rho_1(M^2)$ does not grow any faster than $M^2$ in the ultraviolet.

We can express the operator of Eq.~(\ref{eq.Bportal.NPoperator}) in the Warsaw basis. After employing the equations of motion for the hypercharge gauge boson, we find (see also~\cite{Wells:2015uba}),
 \begin{align}
\frac{\alpha}{M_{\rm IR}^2}\mathcal{O}_{2B}&=-\frac{\alpha}{M_{\rm IR}^2} g'^2  \Bigg[ \mathcal{O}_{H D}+\frac{1}{4} \mathcal{O}_{H \square} \notag\\
& + \left(Y_{q}\left[\mathcal{O}_{H q}^{(1)}\right]_{i i}+Y_{\ell}\left[\mathcal{O}_{H \ell}^{(1)}\right]_{i i}+Y_{u}\left[\mathcal{O}_{H u}\right]_{i i}+Y_{d}\left[\mathcal{O}_{H d}\right]_{i i}+Y_{e}\left[\mathcal{O}_{H e}\right]_{i i}\right)  \notag\\
& +\Big(Y_{q}^{2}\left[\mathcal{O}_{q q}^{(1)}\right]_{i i j j}+Y_{\ell}^{2}\left[\mathcal{O}_{\ell \ell}\right]_{i i j j}+Y_{u}^{2}\left[\mathcal{O}_{u u}\right]_{i i j j}+Y_{d}^{2}\left[\mathcal{O}_{d d}\right]_{i i j j}+Y_{e}^{2}\left[\mathcal{O}_{e e}\right]_{i i j j}  \notag \\
&+2 Y_{q} Y_{\ell}\left[\mathcal{O}_{\ell q}^{(1)}\right]_{i i j j}+2 Y_{q} Y_{u}\left[\mathcal{O}_{q u}^{(1)}\right]_{i i j j}+2 Y_{q} Y_{d}\left[\mathcal{O}_{q d}^{(1)}\right]_{i i j j}+2 Y_{q} Y_{e}\left[\mathcal{O}_{q e}\right]_{i i j j}  \notag\\
&+2 Y_{\ell } Y_{u}\left[\mathcal{O}_{\ell  u}\right]_{i i j j}+2 Y_{\ell } Y_{d}\left[\mathcal{O}_{\ell  d}\right]_{i i j j}+2 Y_{\ell } Y_{e}\left[\mathcal{O}_{\ell  e}\right]_{i i j j} \notag\\
&+2 Y_{u} Y_{d}\left[\mathcal{O}_{u d}^{(1)}\right]_{i i j j}+2 Y_{u} Y_{e}\left[\mathcal{O}_{e u}\right]_{i i j j}+2 Y_{d} Y_{e}\left[\mathcal{O}_{e d}\right]_{i i j j}\Big)\Bigg]\quad {\rm with} \quad \alpha>0 \; .
\label{hyperchargeWarsaw}
 \end{align}
Here $g^\prime$ is the coupling of the SM $U(1)_{Y}$ gauge group, $i$ and $j$ are flavor indices, and $Y_{f}$ denotes the hypercharge of the corresponding fermion $f$, 
\beq
\Big\{Y_q,Y_\ell ,Y_u, Y_d, Y_e\Big\}=\Big\{\frac{1}{6},-\frac{1}{2},\frac{2}{3},-\frac{1}{3},-1\Big\}.
\eeq
The definitions of the Warsaw basis operators $\mathcal{O}_{HD}, \mathcal{O}_{H\Box}, \mathcal{O}_{Hf}$, and $\mathcal{O}_{ff'}$ in this expression are given in Table~\ref{tab:warsaw_basis}.

We now consider the phenomenological implications of the operator in Eq.~(\ref{eq.Bportal.NPoperator}). We see from Eq.~(\ref{hyperchargeWarsaw}) that through the operator $\mathcal{O}_{HD}$, it gives rise to a correction to the mass splitting between the $W$ and $Z$ gauge bosons after the Higgs acquires a VEV. This is very tightly constrained by data. In addition, through the operators of the form $\mathcal{O}_{Hf}$ the couplings of the fermions to the $Z$ gauge boson are altered. This is also very tightly constrained by experiment, as we shall see in Section~\ref{s.constraints}.

\section{Constraints on Portal-Generated SMEFT Operators
\label{s.constraints}}

In the previous section, we determined the leading dimension-six operators that arise from a hidden sector that couples to the SM through the Higgs, neutrino and hypercharge portals. In this section, we consider each of these portals in turn and determine the current constraints on the corresponding dimension-six operators. We parameterize the SMEFT Lagrangian as
\begin{align}
\mathcal{L}_{\mathrm{eff}}= \mathcal{L}_{\rm SM}+\sum_i \frac{C_i}{\Lambda^2} \mathcal{O}_i \;,
\end{align}
where the $\mathcal{O}_i$ denote dimension-six operators in the Warsaw basis  and the $C_i$ represent dimensionless Wilson coefficients. We use EWPO, Higgs measurements, and diboson production data from the LHC to place constraints on the Wilson coefficients of the operators that arise from portal interactions. We make use of the \texttt{HEPfit} package~\cite{DeBlas:2019ehy} to determine the constraints.

In determining the constraints, we take into account the renormalization group evolution from the matching scale $\Lambda$ down to the weak scale. At one-loop order, the Wilson coefficients $C_i$ at the matching scale $\Lambda$ are related to those at the renormalization scale $\mu$ as
 \begin{align}
C_i (\mu) &= C_i (\Lambda) - \frac{\dot C_i}{16\pi^2}\ln\!\bigg(\frac{\Lambda}{\mu}\bigg),	\qquad {\rm with}\qquad \dot C_i\equiv\sum_j \gamma_{ij} C_j \;.
 \end{align}
The anomalous dimension matrix $\gamma_{ij}$ depends on the specific dimension-six SMEFT operators $\mathcal{O}_i$ and $\mathcal{O}_j$. The complete list of one-loop renormalization group equations for the dimension-six SMEFT operators is given in Refs.~\cite{Jenkins:2013zja, Jenkins:2013wua, Alonso:2013hga}.

We employ EWPO data from $W$ and $Z$ pole measurements~\cite{ALEPH:2005ab}. The fit incorporates the following observables: 
\begin{equation}
\alpha, G_\mu, M_Z, M_W, \Gamma_Z, \sigma_h, A_{\ell,\rm{FB}}, A_{b,\rm{FB}}, A_b, A_c, A_\ell, R_\ell, R_b, R_c \;.
\end{equation}
The asymmetries above are defined as
\begin{align}
A_{\ell}&=\frac{\Gamma(Z\to \ell^+_L \ell^-_L)-\Gamma(Z\to \ell^+_R \ell^-_R)}{\Gamma(Z\to \ell^+ \ell^-)},		\notag \\
A_{q}&=\frac{\Gamma(Z\to q_L \bar{q}_L)-\Gamma(Z\to q_R \bar{q}_R)}{\Gamma(Z\to q\bar{q})} \;.
\end{align}
where $\ell = e,\mu,\tau$ and $q=b,c$. 
 The contributions to the $Z$ width are parametrized as
 \begin{align}
R_{\ell}=\frac{\Gamma(Z\to {\rm hadrons})}{\Gamma(Z\to \ell\ell)},\qquad R_{q}=\frac{\Gamma(Z\to q\bar{q})}{\Gamma(Z\to {\rm hadrons})} \;,
 \end{align}
where again $\ell = e,\mu,\tau$ and $q=b,c$.
 The forward-backward symmetries are defined as
\begin{align}
A_{i,\rm{FB}}\equiv \frac{\sigma^i_{\rm F}-\sigma^i_{\rm B}}{\sigma^i_{\rm F}+\sigma^i_{\rm B}}\,,
\end{align}
where $\sigma^i_{\rm F}$ corresponds to the total cross section for angles $\theta$ between the incoming and outgoing particles that lie in the range $(0,\frac{\pi}{2})$ and $\sigma^i_{\rm B}$ to angles between $(\frac{\pi}{2},\pi)$. We incorporate Higgs data from~\cite{Aad:2019mbh,CMS:2020gsy,ATLAS:2022jtk}, $W$ boson mass data from \cite{ALEPH:2013dgf,CDF:2013dpa,ATLAS:2024erm,CMS:2024nau}, and diboson data from~\cite{ALEPH:2013dgf,ATLAS:2017bbg,CMS:2019efc,CMS:2019ppl,ATLAS:2019bsc}.We do not include the recent result for the $W$ boson mass by the CDF collaboration~\cite{CDF:2022hxs}. 

The complete list of the dimension-six Warsaw basis operators generated from each of the three portals is given in \tab{tab:warsaw_basis}. At tree level, the $W$ and $Z$ pole observables are sensitive to the following subset of these operators, see for example~\cite{Dawson:2020oco, Ellis:2020unq},
\begin{align}
\mathcal{O}_{HD}, \mathcal{O}_{Hq}^{(1)}, \mathcal{O}_{H \ell }^{(1)}, \mathcal{O}_{H \ell }^{(3)}, \mathcal{O}_{He}, \mathcal{O}_{Hu}, \mathcal{O}_{Hd}, \mathcal{O}_{\ell \ell }\, .
\end{align}
The corrections to Higgs observables at tree level can arise from the operators
\begin{align}
\mathcal{O}_{H},\mathcal{O}_{H\Box}, \mathcal{O}_{HD}, \mathcal{O}_{Hq}^{(1)}, \mathcal{O}_{H \ell }^{(1)},\mathcal{O}_{H \ell }^{(3)}, \mathcal{O}_{He}, \mathcal{O}_{Hu}, \mathcal{O}_{Hd}, \mathcal{O}_{\ell \ell } \, .
\end{align}
 The diboson $WW$ and $WZ$ data are sensitive to the operators
\begin{align}
\mathcal{O}_{HD},  \mathcal{O}_{Hq}^{(1)}, \mathcal{O}_{H \ell }^{(3)}, \mathcal{O}_{Hu}, \mathcal{O}_{Hd}, \mathcal{O}_{\ell \ell }\, .
\end{align}
In the following subsections, we perform a global fit to the dimension-6 effective operators generated by each of the three portals.


\subsection{The Higgs Portal \label{s.con.higgs}}

As discussed above, after integrating out a hidden sector that couples through the Higgs portal, we generate two independent SMEFT operators $\mathcal{O}_H$ and $\mathcal{O}_{H\Box}$ at the matching scale $\Lambda$ (see \tab{tab:warsaw_basis}).  
The Wilson coefficient $C_{H\Box}$ of the operator $\mathcal{O}_{H\Box}$ at the EFT matching scale $\Lambda$ is related to the parameters in Eq.~\eqref{alphadef} as
\begin{equation}
    \frac{C_{H\Box}}{\Lambda^2} \equiv -\frac{\alpha}{M_{\rm IR}^2}.
\end{equation}
The arguments in the previous section imply that $C_{H\Box}\leq 0$ for
the range of scaling dimensions $\Delta_S \leq 3$.

The immediate effect of the $\mathcal{O}_{H\Box}$ operator is to modify the Higgs kinetic term, resulting in a universal correction to the couplings of the Higgs to the SM fermions and gauge bosons. The corresponding coefficient $C_{H\Box}$ is therefore constrained by LHC data on the Higgs signal strength. In addition, on renormalization group evolution from the matching scale $\Lambda$ down to the weak scale, the operator $\mathcal{O}_{H\Box}$ generates the operators $\mathcal{O}_{HD}$, $\mathcal{O}_{Hq}^{(1)}$, and $\mathcal{O}_{Hq}^{(3)}$ at the weak scale, which are severely constrained by EWPO. The contribution to the operator $\mathcal{O}_{HD}$, related to the oblique $T$-parameter, is the most significant for EWPO. The corresponding renormalization group equation takes the form (see e.g.~\cite{Dawson:2020oco}),
 \begin{align}
\dot{C}_{HD}&=\frac{8}{3}g^{\prime}\Big[2C_{Ht}-C_{Hb}+C_{Hq}^{(1)}\Big]+\frac{20}{3}g^{\prime 2}C_{H\Box}-24\Big[y_t^2C_{Ht}-y_b^2C_{Hb}\Big]+24(y_t^2-y_b^2)C_{Hq}^{(1)} \; ,		
\label{eq_dotCHD}
 \end{align}
where the $y_f$ are Yukawa couplings and we are showing only the most important terms. Until a few years ago, this was the dominant effect driving the limits. However, at present this effect is not as constraining as the direct Higgs measurements.   

The $\mathcal{O}_H$ operator, on the other hand, introduces corrections to the Higgs trilinear and quartic self-interactions. These have not been measured with good precision at the LHC. In addition, renormalization group evolution also does not generate the operators that EWPO are most sensitive to, and so the constraints on $C_{H}$ are comparatively weak.   

\begin{table}[t!]
\centering
\renewcommand{\arraystretch}{1.5}
\begin{tabular}{||c|c||c|c||}
\hline
\multicolumn{4}{||c||}{Higgs portal}\\
\hline
$\mathcal{O}_H$&$(H^\dag H)^3$&$\mathcal{O}_{H\square}$&$\left(H^{\dagger} H\right) \square\left(H^{\dagger} H\right)$ \\
\hline\hline
\multicolumn{4}{||c||}{Neutrino portal}\\
\hline
$\mathcal{O}_{H \ell }^{(1)}$&$(H^{\dagger} i \stackrel{\leftrightarrow}{D}_{\mu} H)(\bar{\ell }_{p} \gamma^{\mu} \ell_{r})$&$\mathcal{O}_{H \ell }^{(3)}$&$(H^{\dagger} i \stackrel{\leftrightarrow}{D}_{\mu}^{I} H)(\bar{\ell }_{p} \tau^{I} \gamma^{\mu} \ell_{r})$ \\
\hline\hline
\multicolumn{4}{||c||}{Hypercharge portal}\\
\hline
$\mathcal{O}_{HD}$&$(H^{\dagger}  {D}_{\mu} H)^\star (H^{\dagger} {D}^{\mu} H)$&$\mathcal{O}_{H\square}$&$\left(H^{\dagger} H\right) \square\left(H^{\dagger} H\right)$ \\
$\mathcal{O}_{Hq}^{(1)}$&$(H^{\dagger} i \stackrel{\leftrightarrow}{D}_{\mu} H)(\bar{q}_{p} \gamma^{\mu} q_{r})$&$\mathcal{O}_{H \ell }^{(1)}$&$(H^{\dagger} i \stackrel{\leftrightarrow}{D}_{\mu} H)(\bar{\ell }_{p} \gamma^{\mu} \ell_{r})$ \\
$\mathcal{O}_{Hu}$&$(H^{\dagger} i \stackrel{\leftrightarrow}{D}_{\mu} H)(\bar{u}_{p} \gamma^{\mu} u_{r})$&$\mathcal{O}_{Hd}$&$(H^{\dagger} i \stackrel{\leftrightarrow}{D}_{\mu} H)(\bar{d}_{p} \gamma^{\mu} d_{r})$ \\
$\mathcal{O}_{He}$&$(H^{\dagger} i \stackrel{\leftrightarrow}{D}_{\mu} H)(\bar{e}_{p}  \gamma^{\mu} e_{r})$&$\mathcal{O}_{ff^\prime}$&{four-fermion operators} \\
\hline
\end{tabular}
\caption{The SMEFT operators generated through the Higgs, neutrino and hypercharge portals in the Warsaw basis.}\label{tab:warsaw_basis}
\end{table}

\begin{figure} [t!]
\centering
\includegraphics[width=0.49\textwidth]{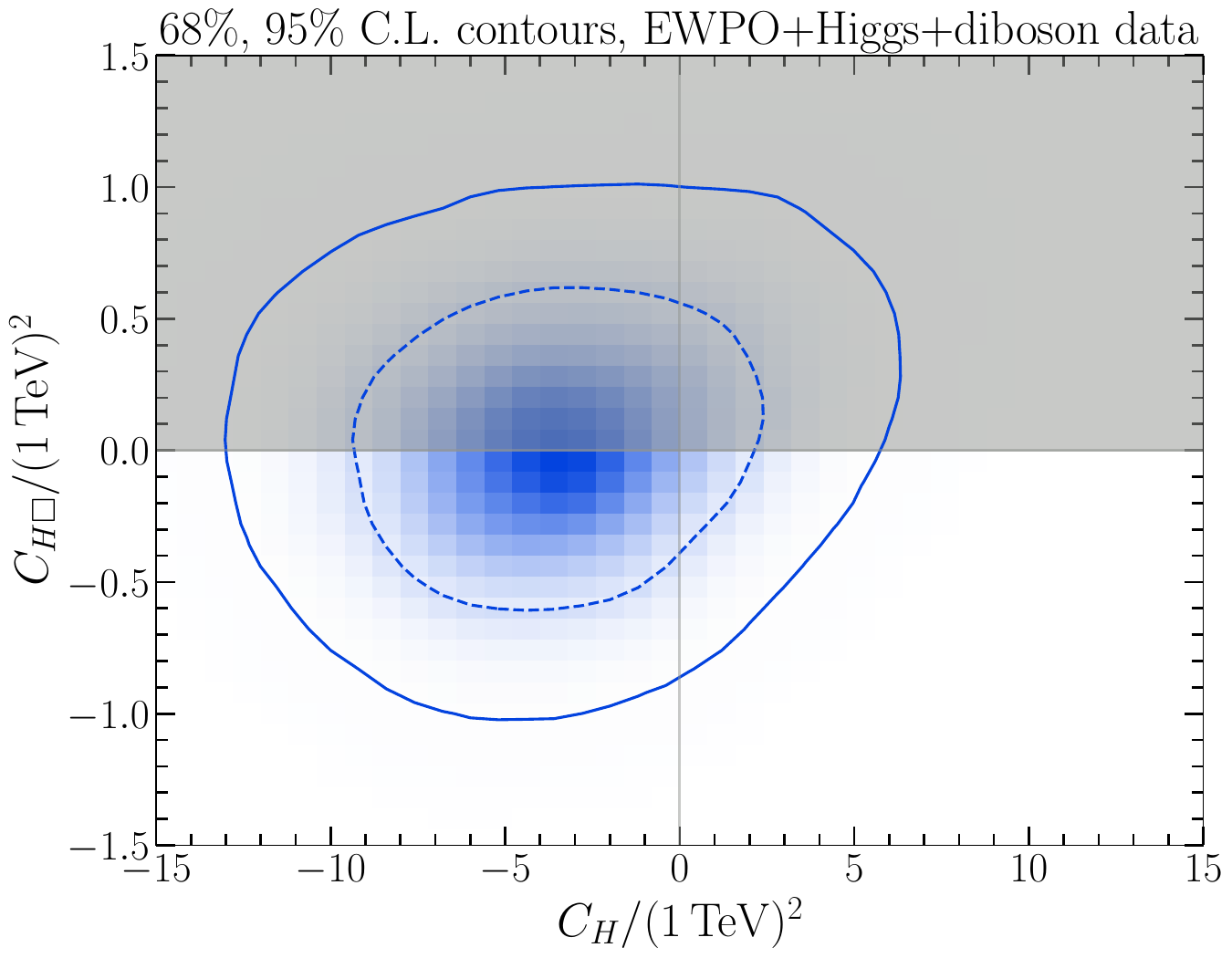}
\includegraphics[width=0.49\textwidth]{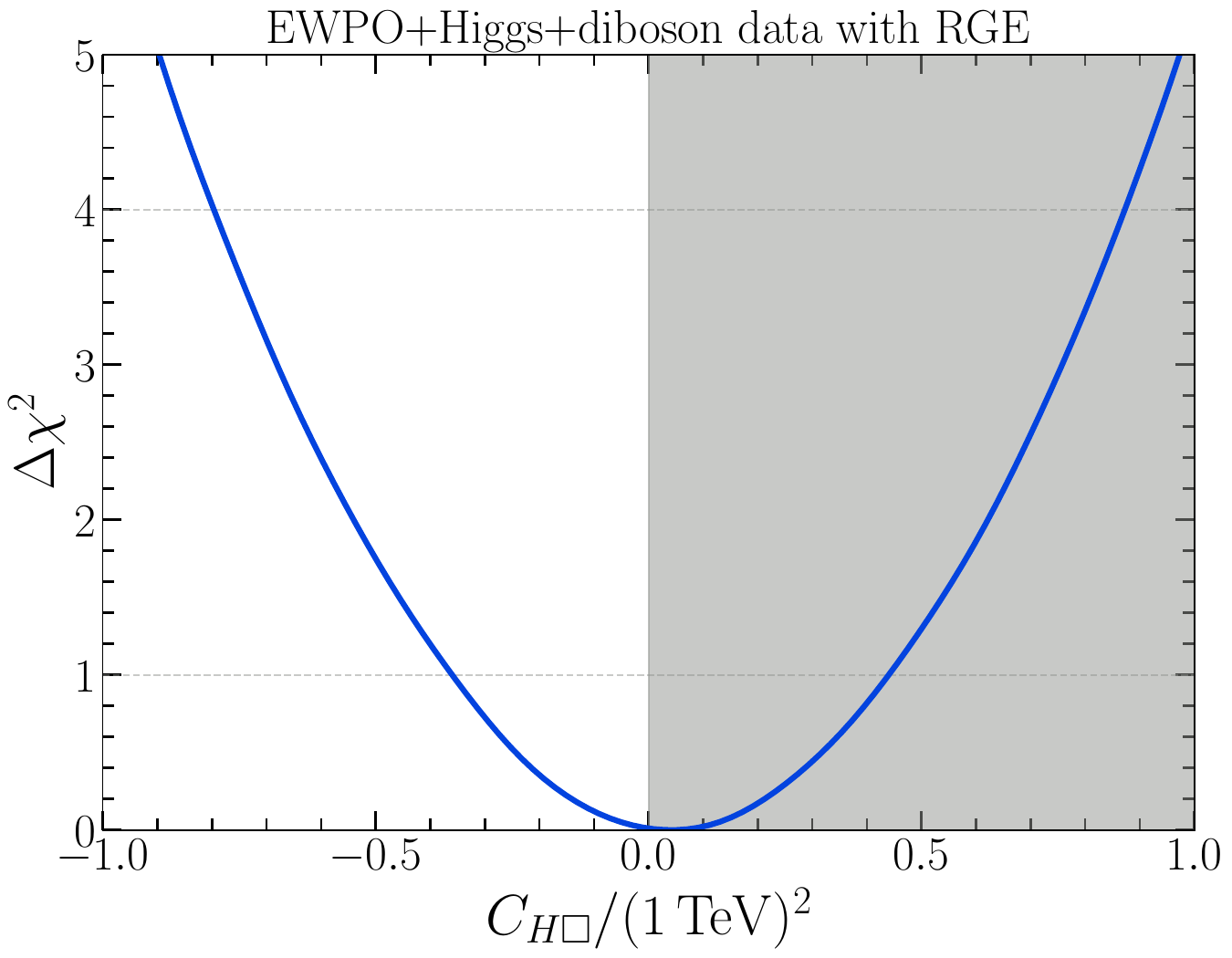}
\caption{The left plot shows the allowed parameter space (1- and 2$\sigma$ contours) for the coefficients $C_H$ and $C_{H\Box}$ of the Higgs portal operators $\mathcal{O}_H$ and $\mathcal{O}_{H\Box}$ for a matching scale $\Lambda=1\tev$. The right plot shows $\Delta \chi^2$ for $C_{H\Box}$ alone for the same matching scale. The gray-shaded regions are forbidden for some range of scaling dimensions of the operator $\mathcal{O}_S$, as discussed in the text. The plots incorporate data from EWPO, Higgs, and diboson observables and take into account renormalization group evolution.}
\label{fig:ffig_CH_CHbox}
\end{figure}

In the left plot of \fig{fig:ffig_CH_CHbox}, we depict the combined 1 and 2-$\sigma$ contours for the Higgs portal operators with Wilson coefficients $C_H$ and $C_{H\Box}$. The matching scale has been taken to be $\Lambda=1\ \text{TeV}$. The gray-shaded region is forbidden for the range of scaling dimensions $\Delta_S \leq 3$, as discussed in \sec{s.th.higgs}.
From the plot, it is evident that $C_{H\Box}$ is much more strongly constrained than $C_H$. In the right plot of \fig{fig:ffig_CH_CHbox}, we show how $\Delta\chi^2$ varies as a function of $C_{H\Box}$. The matching scale has again been taken to be $\Lambda=1\tev$. From the plots, it is clear that the SM point, $C_H = C_{H\Box} = 0$, is in good agreement with the data. 

In these plots, the matching scale has been held fixed at $\Lambda = 1$ TeV. However, we have verified that any value of the matching scale $\Lambda$ in the range from 250 GeV to 10 TeV leads to very similar constraints on the ratios $C_{H\Box}/\Lambda^2$ and $C_{H}/\Lambda^2$. It follows from this that the effects of renormalization group evolution are not significant. 

\subsection{The Neutrino Portal \label{s.con.neutrino}}

In the previous section, we saw that, for a single generation of SM fermions, integrating out a hidden sector coupled through the neutrino portal generates the dimension-six operator $\mathcal{O}_{\ell H}$ \eqref{eq.Nportal.NPoperator} in the SMEFT. We relate the Wilson coefficient of the neutrino portal operator $C_{\ell H}$ at the EFT matching scale $\Lambda$ with the parameters in Eq.~\eqref{eq.Nportal.NPoperator} as,
\begin{equation}
    \frac{C_{\ell H}}{\Lambda^2} \equiv \frac{\alpha}{M_{\rm IR}^2}.
\end{equation}
The arguments in the previous section imply that $C_{\ell H}\geq 0$ for
$\Delta_F \leq 5/2$.
In the Warsaw basis, the neutrino portal operator $\mathcal{O}_{\ell H}$ corresponds to a specific linear combination of the operators $\mathcal{O}_{H\ell}^{(3)}$ and $\mathcal{O}_{H\ell}^{(1)}$ \eqref{eq.NPWarsawbasis}, with the Wilson coefficients $C_{H\ell}^{(1)} = -C_{H\ell}^{(3)}=C_{\ell H}/4$. Generalizing to three generations, we assume that the couplings of the hidden sector to the SM are universal and flavor diagonal, so that $C_{\ell H}$ is the same for all three generations of leptons at the matching scale $\Lambda$.

\begin{figure} [t!]
\centering
\includegraphics[width=0.7\textwidth]{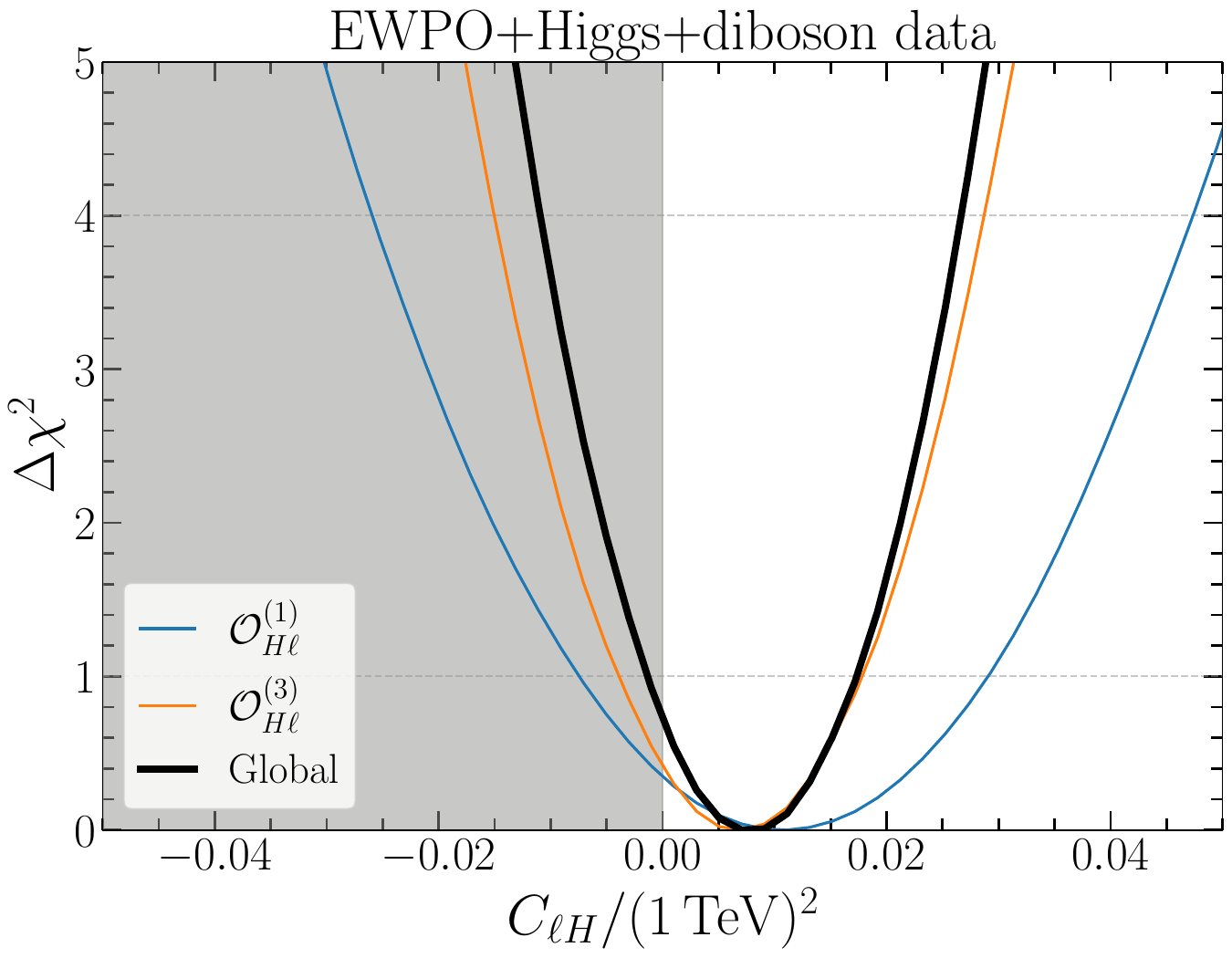}
\caption{We plot $\Delta \chi^2$ as a function of the coefficient $C_{\ell H}$ of the neutrino portal operator $\mathcal{O}_{\ell H}$ (black) for a matching scale $\Lambda=1\tev$, along with the individual contributions from the Warsaw basis operators ${\cal O}_{H\ell}^{(1)}$ and ${\cal O}_{H\ell}^{(3)}$. The gray-shaded region is forbidden for some range of scaling dimensions of the operator $\mathcal{O}_F$, as discussed in the text. The plot incorporates data from EWPO, Higgs, and diboson observables and takes into account renormalization group evolution.}
\label{fig:neutrinoportalchi2}
\end{figure}

In \fig{fig:neutrinoportalchi2}, we plot $\Delta\chi^2$ as a function of the Wilson coefficient of the neutrino portal operator $C_{\ell H}$. We also display the individual contributions to the fit from the SMEFT operators ${\cal O}_{H\ell}^{(1)}$ and ${\cal O}_{H\ell}^{(3)}$. The matching scale $\Lambda$ has been taken as $1\ \text{TeV}$. 
The region forbidden for scaling dimension $\Delta_F \leq 5/2$ has been shaded grey. The fit shows no significant preference for a neutrino portal coupling over the SM. We have verified that the fit is quite insensitive to the value of the matching scale $\Lambda$. In particular,  
for $\Lambda$ anywhere in the range between 250 GeV and 10 TeV, the limits on $C_{\ell H}/\Lambda^2$ change very little, showing that renormalization group effects are not significant.

\subsection{The Hypercharge Portal \label{s.con.hypercharge}}

In the previous section, we saw that integrating out a hidden sector coupled through the hypercharge portal generates the dimension-six operator ${\cal O}_{2B}$ from Eq.~\eqref{eq.Bportal.NPoperator}. In the Warsaw basis, ${\cal O}_{2B}$ corresponds to a specific linear combination of several different operators in the SMEFT as described in Eq.~\eqref{hyperchargeWarsaw}. The global fit is most sensitive to the operators
$\mathcal{O}_{H \ell }^{(1)}=(H^{\dagger} i \stackrel{\leftrightarrow}{D}_{\mu} H)(\bar{l}_{p} \gamma^{\mu} l_{r})$, $\mathcal{O}_{H e} =(H^{\dagger} i \stackrel{\leftrightarrow}{D}_{\mu} H)(\bar{e}_{p}  \gamma^{\mu} e_{r})$ and $\mathcal{O}_{H D}=(H^{\dagger}  {D}_{\mu} H)^* (H^{\dagger} {D}_{\mu} H)$, which are severely constrained by EWPO. 

\begin{figure} [t!]
\centering
\includegraphics[width=0.7\textwidth]{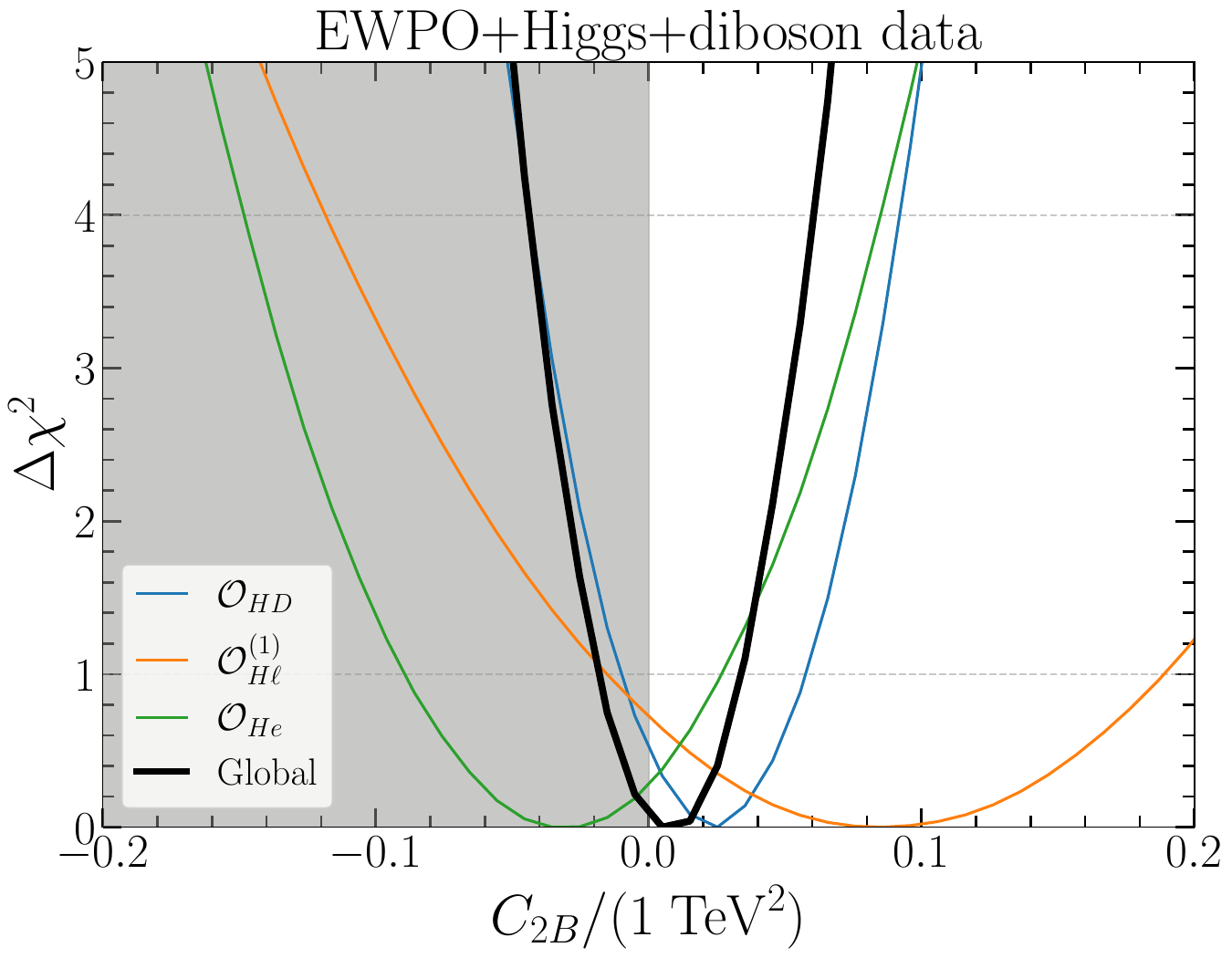}
\caption{We plot $\Delta \chi^2$ as a function of the coefficient $C_{2B}$ of the hypercharge portal operator $\mathcal{O}_{2B}$ (black) for a matching scale $\Lambda=1\tev$, along with the contributions of the individual Warsaw basis operators. The gray-shaded region is forbidden for some range of scaling dimensions of the operator $\mathcal{O}_T$, as discussed in the text. The plot incorporates data from EWPO, Higgs, and diboson observables and takes into account renormalization group evolution.}
\label{fig:C2BwithRGE}
\end{figure}

In \fig{fig:C2BwithRGE}, we present the results of our global fit expressed in terms of the Wilson coefficient $C_{2B}$ of the ${\cal O}_{2B}$ operator. This is related to the parameters in \eqref{eq.Bportal.NPoperator} as
\begin{equation}
    \frac{C_{2B}}{\Lambda^2} \equiv \frac{\alpha}{M_{\rm IR}^2} \;.
\end{equation}
The arguments in the previous section imply that $C_{2B}\geq 0$ for the range of scaling dimensions $\Delta_T \leq 3$. In the figure, we plot $\Delta\chi^2$ as a function of the Wilson coefficient $C_{2B}$ of the operator ${\cal O}_{2B}$. We also display the individual contributions to 
$\Delta\chi^2$ from the most constraining Warsaw basis operators. We have again taken the matching scale $\Lambda$ to be $1\tev$ and included the effects of renormalization group evolution on the Wilson coefficients. The region forbidden for scaling dimension $\Delta_T \leq 3$ has been shaded gray. We have verified that for matching scales $\Lambda$ anywhere in the range between 250 GeV and 10 TeV, the results of the global fit change very little, showing that renormalization group effects are not significant. The fit shows no preference for a hidden sector coupled through the hypercharge portal over the SM. Near-future LHC measurements are expected to have much greater sensitivity to $C_{2B}$~\cite{Torre:2020aiz}. 
For the numerical study, we have employed the \texttt{HEPfit} package~\cite{DeBlas:2019ehy}, where the ${\cal O}_{2B}$ operator is implemented (with a sign convention opposite to ours in Eq.~\eqref{eq.Bportal.NPoperator}) in terms of Warsaw basis operators Eq.~\eqref{hyperchargeWarsaw}. 
To ensure the robustness of our results, we have cross-checked them using a modified version of the \texttt{Fitmaker} package~\cite{Ellis:2020unq}, and found good agreement.

\section{Summary and Outlook \label{s.summary}}

In this paper, we have explored universal features of the effect of hidden sectors on the SM. We have considered hidden sectors that couple to the SM through one of the lowest dimension portal interactions, namely the Higgs, neutrino, or hypercharge portals, and determined the forms of the leading dimension-six terms that are generated in the low energy effective theory 
when the hidden sector is integrated out. Our results show that, for any specific portal interaction, the forms of the leading dimension-six operators are fixed and independent of details of the hidden sector.

In the case of the Higgs portal, we find that two independent dimension-six terms are generated, one of which has a sign that, under certain conditions, is restricted by the requirement that the dynamics in the hidden sector be causal and unitary. In the case of the neutrino portal, for a single generation of SM fermions and assuming that the hidden sector does not
violate lepton number, a unique dimension-six operator is generated. For the hypercharge portal, again a
unique dimension-six operator is generated. For both the neutrino and hypercharge portals, under certain conditions, the signs of the coefficients of these operators are fixed by the requirement that the hidden sector be causal and unitary. Our results for the general forms of the operators and their signs are in agreement with the results for the corresponding portals in the minimal hidden sector models considered in Ref.~\cite{deBlas:2017xtg}.

\begin{figure} [t!]
\centering
\includegraphics[width=0.7\textwidth]{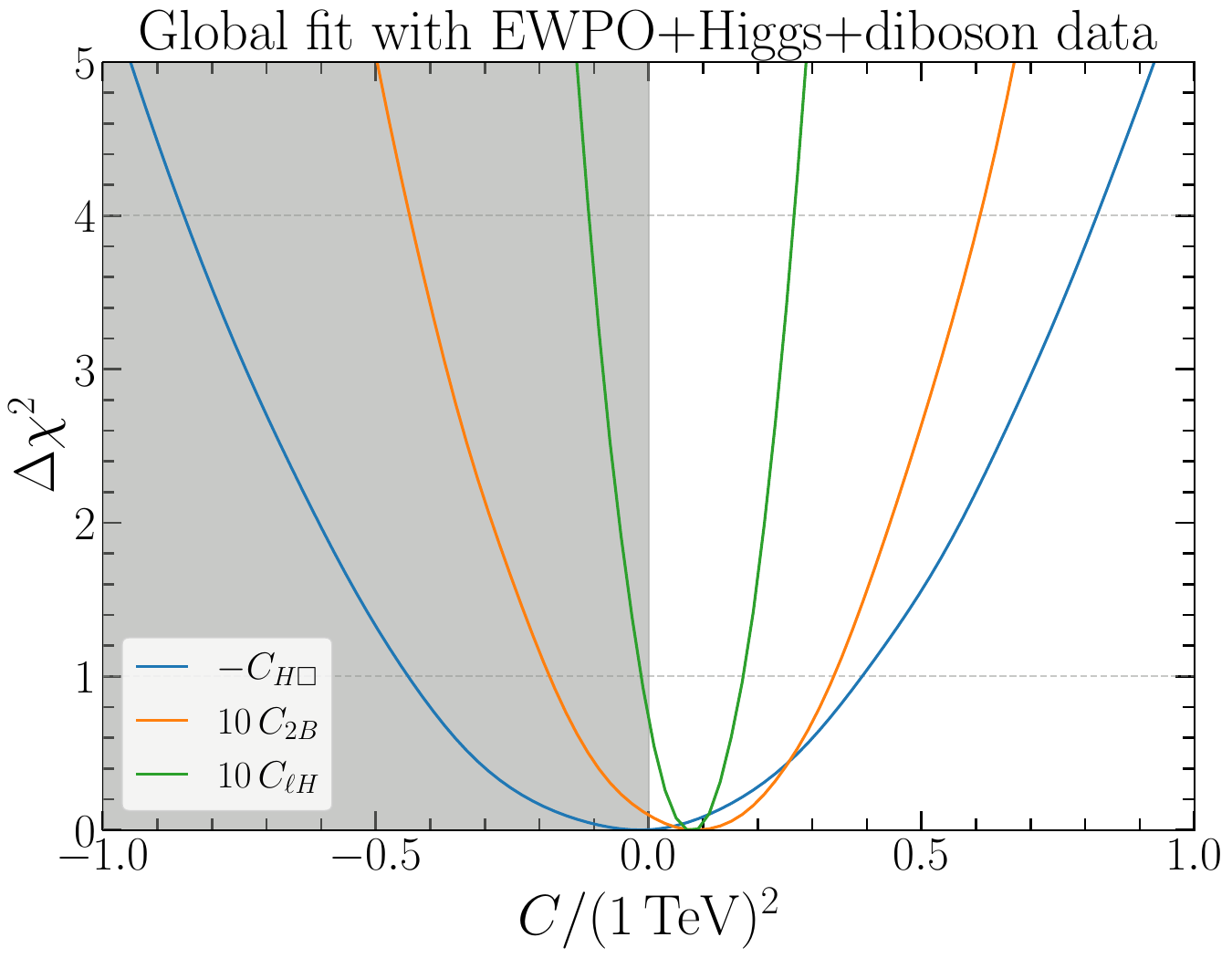}
\caption{We plot $\Delta \chi^2$ as a function of Wilson coefficients of the Higgs, neutrino, and hypercharge portal operators for a matching scale $\Lambda=1\tev$. The plot incorporates data from EWPO, Higgs, and diboson observables and takes into account renormalization group evolution. The gray-shaded region is forbidden for some range of scaling dimensions of the hidden sector operator $\mathcal{O}_{\rm HS}$, as discussed above.}
\label{fig:global}
\end{figure}

For each portal, we have discussed the experimental implications of the leading operators. Translating the operators in question into the Warsaw basis, we have numerically evaluated the constraints on them from EWPO, Higgs measurements, and diboson production data. We find that there is no significant preference for a hidden sector coupled via the Higgs, neutrino or hypercharge portals over the SM. In Fig.~\ref{fig:global} we present a summary plot showing $\Delta \chi^2$ as a function of Wilson coefficients of the Higgs, neutrino, and hypercharge portal operators for a matching scale $\Lambda=1\tev$. In this plot, the shaded region is excluded for some range of scaling dimensions by the positivity considerations discussed above, and we have employed flat priors for the Monte-Carlo analysis.


\section*{Acknowledgements}
We thank Lorenzo Ricci for insightful discussions. ZC and IF are supported by the National Science Foundation under Grant Number PHY-2210361. The research of CK is supported by the National Science Foundation under Grant Number PHY-2210562. 

\appendix
\section{Regulating the Dispersive Integral}
\label{appendix1}

In performing the analysis leading up to Eq.~(\ref{eq.Hportal.NPoperator}), we implicitly assumed that the integral in Eq.~(\ref{integral}) does not diverge in the ultraviolet. In general, this assumption may not be valid and must be reexamined. We begin by considering the case of an operator $\mathcal{O}_S$ that has a definite scaling dimension $\Delta_S$ in the ultraviolet, so that $\rho(M^2)$ scales as $(M^2)^{\Delta_S - 2}$ for large $M^2$. Then the integral in Eq.~(\ref{integral}) diverges in the ultraviolet for
$\Delta_S \geq 2$ and must be regulated. In this appendix, we discuss the procedure for regulating this integral and its relation to conventional renormalization schemes. It is convenient to define
 \begin{equation}
\Pi_F(p^2) = \int_0^\infty dM^2 \rho(M^2)
             \frac{i}{p^2 - M^2 + i \epsilon} \;.
 \end{equation}
 For $2 \leq \Delta_S < 3$, we may regulate this by noting that the difference,
 \begin{equation}
 \label{georgi-norm-reg1}
\Pi_F(p^2) - \Pi_F(p_0^2) =
\int_0^\infty dM^2 \rho(M^2)
\frac{i(p_0^2 - p^2)}
{\left(p^2 - M^2 + i \epsilon\right)
 \left(p_0^2 - M^2 + i \epsilon \right)} \; ,
 \end{equation}
 is finite in the ultraviolet. Here $\Pi_F(p_0^2)$ is the value of $\Pi_F(p^2)$ at some reference off-shell momentum $p_0^2 < 0$.  We can rewrite this as
 \begin{equation}
 \label{georgi-norm-reg2}
\Pi_F(p^2) = \Pi_F(p_0^2) +
\int_0^\infty dM^2 \rho(M^2)
\frac{i(p_0^2 - p^2)}
{\left(p^2 - M^2 + i \epsilon\right)
 \left(p_0^2 - M^2 + i \epsilon \right)} \; ,
 \end{equation}
 where, in this expression, $\Pi_F(p_0^2)$ is to be treated as a free 
parameter that incorporates the unknown ultraviolet physics. It is closely related to the counterterms in a conventional renormalization scheme. Since 
$p_0^2 < 0$, the absorptive part of Eq.~(\ref{georgi-norm-reg2}) is 
exactly the same as that of Eq.~(\ref{integral}).

 For $\Delta_S \geq 2$, the Higgs portal interaction in 
Eq.~(\ref{HPortal1}) gives rise to a divergent contribution to the 
amplitude for the scattering of Higgs fields, $H H \rightarrow H H$ at 
order ${\lambda}^2$. The corresponding matrix element is of the schematic 
form,
 \begin{equation}
\label{matrix-element}
\langle p_3, p_4 | 
\int d^4 x H^{\dagger}(x) H(x) \mathcal{O}_S(x) 
\int d^4 y H^{\dagger}(y) H(y) \mathcal{O}_S(y) 
| p_1, p_2 \rangle \; ,
 \end{equation}
 which scales as 
 \begin{equation}
\Pi_F([p_1 - p_3]^2) + \Pi_F([p_1 - p_4]^2)\,.
 \end{equation}
 Once $\Pi_F(p^2)$ has been regulated by the procedure in Eqs.~(\ref{georgi-norm-reg1}) and 
(\ref{georgi-norm-reg2}), this matrix element is also finite. 

The procedure we have outlined is equivalent to a conventional renormalization scheme such as momentum subtraction. To illustrate this, we write the Higgs quartic term as the sum of a renormalized parameter and a counterterm,
 \begin{equation}
- \mathcal{L} \supset \lambda_H (H^{\dagger} H)^2 +
                      \delta_{\lambda_H} (H^{\dagger} H)^2 \; ,
 \end{equation}
 where the value of the renormalized parameter $\lambda_H$ is set by the 
scattering amplitude at some reference momenta ($\hat{p}_1,\hat{p}_2;\hat{p}_3,\hat{p}_4$) chosen such that $(\hat{p}_1 - \hat{p}_3)^2 = (\hat{p}_1 - \hat{p}_4)^2 = p_0^2$, where $|p_0^2| \gg M_{\rm IR}^2$. With this definition, the counterterm $\delta \lambda_H$ is given by
 \begin{equation}
\delta_{\lambda_H} = 2 i \lambda^2 \Pi_F(p_0^2) \; ,
 \end{equation}
 where, for simplicity, we are neglecting the SM contributions. Then the sum of the (regulated) contributions to the amplitude from 
the interaction in Eq.~(\ref{HPortal1}) cancel exactly 
against the counterterm contribution when the incoming and outgoing particles have the reference momenta, but sum to a 
non-zero value for other values of the external momenta. For general incoming and outgoing momenta, the matrix element is given by
 \begin{align}
 -i\lambda_H - \lambda^2\left\{\Pi_F([p_1-p_3]^2) - \Pi_F(p_0^2)\right\}
 - \lambda^2\left\{\Pi_F([p_1-p_4]^2) - \Pi_F(p_0^2)\right\}\,.
 \end{align}
 This is finite and well-behaved in the ultraviolet. For $|(p_1-p_3)^2| \ll M_{\rm IR}^2$, we can expand out 
 \begin{align}
\left\{\Pi_F([p_1-p_3]^2) - \Pi_F(p_0^2)\right\} =
\left\{\Pi_F(0) - \Pi_F(p_0^2)\right\} -
i\int_0^{\infty}dM^2 \frac{\rho(M^2)}{M^2}\frac{(p_1-p_3)^2}{M^2} \;,
 \label{expandedout}
 \end{align}
 and similarly for $|(p_1 -p_4)^2| \ll M_{\rm IR}^2$. Comparing to Eq.~(\ref{expanded_sdf}), we see that the terms of order $p^2/M^2$ are unchanged by the regulation procedure. It follows from this that the coefficient of the operator $\mathcal{O}_{H\Box}$ is finite and unaltered by the regulation procedure for this range of scaling dimensions, and the prediction for its sign is therefore valid.
 
 In the more general case, the operator $\mathcal{O}_S$ may not have a definite scaling dimension. However, it is clear that the same regularization procedure can be applied provided that $\rho(M^2)$ does not grow any faster than $M^2$ in the ultraviolet. Therefore, if this condition holds, the prediction for the sign of the coefficient of the operator $\mathcal{O}_{H\Box}$ is valid. 

 For $\Delta_S \geq 3$, the right hand side of 
Eq.~(\ref{georgi-norm-reg2}) is still ultraviolet divergent after the renormalization procedure outlined above and additional 
regulation is needed. For $3 \leq \Delta_S < 4$, we note that the 
difference
  \begin{equation}
 \label{georgi-norm-reg3}
\left\{\Pi_F(p^2) - \Pi_F(p_0^2)
- \Pi_F'(p_0^2)\left[p^2 - p_0^2\right]\right\} =
\int_0^\infty dM^2 \rho(M^2)
\frac{i(p_0^2 - p^2)^2}
{\left(p^2 - M^2 + i \epsilon\right)
 \left(p_0^2 - M^2 + i \epsilon \right)^2}
 \end{equation}
 is finite. This can be rewritten as an expression for 
$\Pi_F(p^2)$,
   \begin{equation}
 \label{georgi-norm-reg4}
\Pi_F(p^2) = \Pi_F(p_0^2)
+ \Pi_F'(p_0^2)\left[p^2 - p_0^2\right] +
\int_0^\infty dM^2 \rho(M^2)
\frac{i(p_0^2 - p^2)^2}
{\left(p^2 - M^2 + i \epsilon\right)
 \left(p_0^2 - M^2 + i \epsilon \right)^2} \; .
 \end{equation}
 Here $\Pi_F(p_0^2)$ and $\Pi_F'(p_0^2)$ are to be 
treated as free parameters that incorporate the unknown 
ultraviolet physics. They are related to the counterterms in a conventional renormalization scheme. Once again the absorptive part of 
Eq.~(\ref{integral}) is unaffected by the regularization procedure. 

Once $\Pi_F(p^2)$ has been regulated following the procedure in Eqs.~(\ref{georgi-norm-reg3}) and (\ref{georgi-norm-reg4}), the contribution to the matrix element in Eq.~(\ref{matrix-element}) is finite. In a renormalization scheme based on momentum subtraction, we will now require counterterms corresponding to both $(H^{\dagger} H)^2$ and $(H^{\dagger} H)\partial^2(H^{\dagger} H)$. These are related to the parameters $\Pi_F(p_0^2)$ and $\Pi_F'(p_0^2)$. Since a counterterm is now required for the operator $(H^{\dagger} H)\partial^2(H^{\dagger} H)$, the prediction for the sign of its coefficient is no longer valid, except for the special case of $\Delta_S = 3$, when the remaining divergence in Eq.~(\ref{expandedout}) is only logarithmic.   

\section{Effect of $\mathcal{CPT}$ Invariance on the Spectral Decomposition}
\label{appendix2}

In this appendix, we consider the implications of $\mathcal{CPT}$ invariance for the form of the spectral decomposition of the time-ordered two-point function, 
$\langle \Omega | T\{\mathcal{O}_{\rm HS}^{(\dagger)}(x) \mathcal{O}_{\rm HS}(y)\} |\Omega \rangle$. In general, this consists of a sum of two terms, each corresponding to a different time-ordering. We show that the coefficients of these terms are related by $\mathcal{CPT}$ symmetry and are therefore not independent. We will limit our discussion to the fermion and tensor two-point functions, the scalar case being straightforward.

We first consider the case of the fermion two-point function. From $\mathcal{CPT}$ invariance, 
\begin{align}
    \bra{\Omega}{\mathcal{O}_F(y)\mathcal{O}_F}^\dagger(x)\ket{\Omega} &= \braket{U_\mathcal{CPT}\Omega|U_\mathcal{CPT}\mathcal{O}_F(y)\mathcal{O}_F^\dagger(x)\Omega}^*  \nonumber\\
    &= \braket{U_\mathcal{CPT}\mathcal{O}_F(y)\mathcal{O}_F^\dagger(x)\Omega|U_\mathcal{CPT}\Omega} \;,
\end{align}
where $U_\mathcal{CPT}$ represents the anti-unitary operator that generates the $\mathcal{CPT}$ transformation.
Using the $\mathcal{CPT}$ invariance of the vacuum state 
\begin{equation}
    U_\mathcal{CPT}\ket{\Omega}=\ket{\Omega} \; ,
\end{equation}
we then have
\begin{equation}
    \bra{\Omega}{\mathcal{O}_F(y)\mathcal{O}_F}^\dagger(x)\ket{\Omega}= \braket{U_\mathcal{CPT}\mathcal{O}_F(y)U_\mathcal{CPT}^{-1}U_\mathcal{CPT}\mathcal{O}_F^\dagger(x)U_\mathcal{CPT}^{-1}\Omega|\Omega} \; .
\end{equation}
Under $\mathcal{CPT}$ the spinors $\mathcal{O}_F$ and $\mathcal{O}_F^{\dagger}$ transform as\footnote{For the purposes of this discussion, we follow the conventions of Peskin and Schroeder~\cite{Peskin:1995ev} for the operations $\mathcal{C}$, $\mathcal{P}$ and $\mathcal{T}$, with the arbitrary phase factor $\eta_a$ set to $i$.} 
\begin{align}
    U_\mathcal{CPT} \mathcal{O}_F^{\dagger\dot{\alpha}}(x) U_\mathcal{CPT}^{-1} &= i\mathcal{O}_F^{\alpha}(-x) \\
    U_\mathcal{CPT}\mathcal{O}_F^{\alpha}(y)U_\mathcal{CPT}^{-1} &= -i\mathcal{O}_F^{\dagger\dot{\alpha}}(-y) \; .
\end{align}
 We therefore have
\begin{align}
     \bra{\Omega}{\mathcal{O}_F^\alpha(y)\mathcal{O}_F^{\dagger \dot{\alpha}}(x)}\ket{\Omega} &= \braket{\mathcal{O}_F^{\dagger\dot{\alpha}}(-y)\mathcal{O}_F^{\alpha}(-x) \Omega|\Omega} \nonumber \\
     &= \bra{\Omega} \left(\mathcal{O}_F^{\dagger\dot{\alpha}}(-y)\mathcal{O}_F^{\alpha}(-x)\right)^\dagger \ket{\Omega} \nonumber \\
     &= \bra{\Omega}\mathcal{O}_F^{\dagger\dot{\alpha}}(-x)\mathcal{O}_F^{\alpha}(-y)\ket{\Omega}  \; .
\end{align}
Using this result and the translational invariance of the theory, we can obtain a relation between the coefficients of the two terms in the spectral decomposition of the time-ordered two-point function.

We now turn our attention to the case of the tensor two-point function. From $\mathcal{CPT}$ invariance,  
\begin{align}
    \bra{\Omega}{\mathcal{O}_T^{\alpha\beta}(y)\mathcal{O}_T}^{\mu\nu}(x)\ket{\Omega} &= \braket{U_\mathcal{CPT}\Omega |  U_\mathcal{CPT}\mathcal{O}_T^{\alpha\beta}(y)\mathcal{O}_T^{\mu\nu}(x)\Omega}^* \nonumber \\
    &= \braket{U_\mathcal{CPT}\mathcal{O}_T^{\alpha\beta}(y)\mathcal{O}_T^{\mu\nu}(x)\Omega|U_\mathcal{CPT}\Omega} \nonumber \\
    &=\braket{U_\mathcal{CPT}\mathcal{O}_T^{\alpha\beta}(y)U_\mathcal{CPT}^{-1}U_\mathcal{CPT}\mathcal{O}_T^{\mu\nu}(x)U_\mathcal{CPT}^{-1}\Omega|\Omega} \; .
\end{align}
Under $\mathcal{CPT}$ the tensor $\mathcal{O}_T$ transforms as
\begin{align}
    U_\mathcal{CPT} \mathcal{O}_T^{\mu\nu}(x) U_\mathcal{CPT}^{-1} &=  \mathcal{O}_T^{\mu\nu}(-x) \; .
\end{align}
It follows that
\begin{align}
    \bra{\Omega}{\mathcal{O}_T^{\alpha\beta}(y)\mathcal{O}_T}^{\mu\nu}(x)\ket{\Omega}  &= \braket{\mathcal{O}_T^{\alpha\beta}(-y)\mathcal{O}_T^{\mu\nu}(-x) \Omega|\Omega} \nonumber \\
     &= \bra{\Omega} \left(\mathcal{O}_T^{\alpha\beta}(-y)\mathcal{O}_T^{\mu\nu}(-x)\right)^\dagger \ket{\Omega} \nonumber \\
     &= \bra{\Omega}\mathcal{O}_T^{\mu\nu}(-x)\mathcal{O}_T^{\alpha\beta}(-y)\ket{\Omega}.  
\end{align}
 Using this result and the translational invariance of the theory, we can again obtain a relation between the coefficients of the two terms in the spectral decomposition of the time-ordered two-point function.

\bibliography{bib_globalfit}{}
\bibliographystyle{aabib}

\end{document}